\documentclass[a4paper,reqno]{amsart}

\usepackage{mathrsfs,url,amscd,amssymb}

\usepackage[all]{xy}

\SelectTips{cm}{}

\allowdisplaybreaks[4]

\newcommand*{\pd}[2]{\mathchoice{\frac{\partial#1}{\partial#2}}
  {\partial#1/\partial#2}{\partial#1/\partial#2}
  {\partial#1/\partial#2}}
\newcommand*{\fd}[2]{\mathchoice{\frac{\delta#1}{\delta#2}} {\delta
    #1/\delta#2}{\delta#1/\delta#2}{\delta#1/\delta#2}}

\newcommand{\ldb}{\mathrm{[\![}}
\newcommand{\rdb}{\mathrm{]\!]}}

\newcommand{\enVert}[2][\right]{\relax \ifx#1\right\relax
\left\lVert\else#1\lVert\fi#2#1\rVert} 

\newcommand{\envert}[2][\right]{\relax
 \ifx#1\right\relax \left\lvert\else#1\lvert\fi#2#1\rvert}
\let\abs=\envert

\let\phi\varphi
\let\kappa\varkappa

\newcommand{\cprime}{\/{\mathsurround=0pt$'$}}
\newcommand{\ol}{\bar}

\DeclareFontFamily{OML}{cyi}{} \DeclareFontShape{OML}{cyi}{m}{n}{ <5>
  <6> <7> <8> <9> gen * wncyi <10> <10.95> <12> <14.4> <17.28> <20.74>
  <24.88> wncyi10 }{} \DeclareSymbolFont{rusletters}{OML}{cyi}{m}{n}
\DeclareSymbolFontAlphabet{\rusmath}{rusletters}
\DeclareMathSymbol\re{\rusmath}{rusletters}{"03}

\DeclareMathOperator{\sym}{sym}
\DeclareMathOperator{\Rec}{Rec}

\providecommand{\href}[2]{#2} \providecommand{\urlprefix}{URL }
\providecommand*{\eprint}[2][]{%
\href{http://arXiv.org/abs/#2}{\begingroup \Url{arXiv:#2}}%
}

\newtheorem{theorem}{Theorem}
\newtheorem{proposition}{Proposition}

\theoremstyle{definition}
\newtheorem{example}{Example}

\theoremstyle{remark}
\newtheorem{remark}{Remark}

\begin{document}

\hfill nlin.SI/0511012

\bigskip

\title[The D-Boussinesq Equation]{A geometric study of the
  dispersionless Boussinesq type equation}
\thanks{This work was supported in part by the NWO grant 047017015}
\author[P.~Kersten]{P.~Kersten}

\address{Paul Kersten \\
  University of Twente,  Faculty of Mathematical Sciences \\
  P.O.~Box 217 \\
  7500 AE Enschede \\
  The Netherlands}

\email{kersten@math.utwente.nl}

\author{I.~Krasil{\cprime}shchik}

\address{Iosif Krasil{\cprime}shchik \\
  Independent University of Moscow \\
  B. Vlasevsky~11 \\
  119002 Moscow \\
  Russia}

\email{josephk@diffiety.ac.ru}

\author{A.~Verbovetsky}

\address{Alexander Verbovetsky \\
  Independent University of Moscow \\
  B. Vlasevsky~11 \\
  119002 Moscow \\
  Russia}

\email{verbovet@mccme.ru}



\begin{abstract}
  We discuss the dispersionless Boussinesq type equation, which is
  equivalent to the Benney--Lax equation, being a system of equations
  of hydrodynamical type. This equation was discussed
  in~\cite{GumNut}.The results include: a description of local and
  nonlocal Hamiltonian and symplectic structures, hierarchies of
  symmetries, hierarchies of conservation laws, recursion operators
  for symmetries and generating functions of conservation laws
  (cosymmetries). Highly interesting are the appearances of operators
  that send conservation laws and symmetries to each other but are
  neither Hamiltonian, nor symplectic. These operators give rise to a
  noncommutative infinite-dimensional algebra of recursion operators.
\end{abstract}
\maketitle


\section*{Introduction}

Below we deal with the \emph{dispersionless Boussinesq type equation}
(the dB-equation), which is the system
\begin{align}\label{eq:bouss-new:1}
  w_t&=u_x,\nonumber\\
  u_t&=ww_x+v_x,\\
  v_t&=-uw_x-3wu_x,\nonumber
\end{align}
being equivalent to the Benney--Lax equation, and which is known to be
integrable,~\cite{GumNut}. In particular, it possesses a
bi-Hamiltonian structure. System~\eqref{eq:bouss-new:1} is of
hydrodynamical type and can be obtained as a reduction of the
Khokhlov--Zabolotskaya equation.

Using the methods developed
in~\cite{KerstenKrasilshchikVerbovetsky:HOpC}, we rediscover the above
mentioned bi-Hamiltonian structure and show that it is only a part of
the infinite-dimensional space of operators that take conservation laws
of~\eqref{eq:bouss-new:1} (their generating functions, more
precisely) to symmetries. These operators, in a standard way, generate
an infinite associative (but not commutative) algebra of recursion
operators for symmetries.

Every recursion operator, applied to a known symmetry (e.g., a point
one), gives rise to an infinite family of symmetries (both local and
nonlocal ones). Contrary to the known examples, these families are not
\emph{hierarchies} in the usual sense, because their jet order does
not grow infinitely but remains at level~$1$ for all symmetries we
found.

Dually, there exists an infinite-dimensional space of operators that
take symmetries of the dB-equation to generating functions (or
\emph{cosymmetries}) and only some of these operators determine
symplectic (or inverse Noether) structures on the equation. In a
similar way, we obtain an infinite algebra of recursion operators for
cosymmetries and infinite families of conservation laws (also of $1$st
order).

Below we present a detailed analysis of all these structures. In
Section~\ref{sec:background} a very informal introduction to the
theoretical background is given. Section~\ref{sec:prep-comp} contains
preparatory material needed to achieve the main results. These results
are exposed in Section~\ref{sec:main-results} and discussed in
concluding remarks (Sections~\ref{sec:interrelations}
and~\ref{sec:discussion}).

\section{Background}
\label{sec:background}

As it was mentioned in the Introduction, our computations are based on
the results of paper~\cite{KerstenKrasilshchikVerbovetsky:HOpC} (see
also~\cite{KerstenKrasilshchikVerbovetsky:N=2}). For the general
theoretical background we also refer the reader to
books~\cite{KrasilshchikVinogradov:SCLDEqMP,KrasilshchikKersten:SROpCSDE,KrasilshchikVerbovetsky:HMEqMP}.
Here we shall give an informal description of the computational scheme
we use in subsequent sections.

We consider a system~$\mathcal{E}$ of evolution equations
\begin{equation}
  \label{eq:bouss-new:2}
  u_t=F(x,t,u,u_1,\dots,u_k),
\end{equation}
where both $u=(u^1,\dots,u^m)$ and $F=(F^1,\dots,F^m)$ are vectors
and~$u_t=\pd{u}{t}$, $u_s=\pd{^su}{x^s}$. For simplicity, we restrict
ourselves to the case of one-dimensional space variable~$x$, though
everything works in the general situation as well. We are interested
in \emph{symmetries} and \emph{conservation laws} of
system~\eqref{eq:bouss-new:2} and in various operators that relate
these objects to each other (recursion operators, Hamiltonian and
symplectic structures).

\subsection{Symmetries and conservation laws}
\label{subsec:symm-cons-laws}

A \emph{symmetry} of equation~\eqref{eq:bouss-new:2} is a vector
field (an \emph{evolutionary} field)
\begin{equation}
  \label{eq:bouss-new:3}
  \re_\phi=\sum_{i\ge0}\sum_{j=1}^m
  D_x^i(\phi^j)\pd{}{u_i^j},
\end{equation}
where $\phi=(\phi^1,\dots,\phi^m)$ is a vector function depending on
$x$, $t$, $u$, $u_1,\dots,u_s$ and satisfying the equations
\begin{equation}
  \label{eq:bouss-new:4}
  D_t(\phi^j)=\sum_{i,l}\pd{F^j}{u_i^l}D_x^i(\phi^l),\qquad
  j=1,\dots,m.
\end{equation}
Here and below
\begin{equation*}
  D_x=\pd{}{x}+\sum_{i,l}u_{i+1}^l\pd{}{u_i^l},\qquad
  D_t=\pd{}{t}+\sum_{i,l}D_x^i(F^l)\pd{}{u_i^l}
\end{equation*}
are \emph{total derivatives} with respect to~$x$ and~$t$. We identify
fields~$\re_\phi$ with functions~$\phi$ (the \emph{generating
  functions}). Thus, a symmetry is a function that satisfies the
equation
\begin{equation}
  \label{eq:bouss-new:5}
  \ell_{\mathcal{E}}(\phi)=0,
\end{equation}
where
\begin{equation}
  \label{eq:bouss-new:6}
  \ell_{\mathcal{E}}=D_t-\ell_F
\end{equation}
and
\begin{equation}
  \label{eq:bouss-new:7}
  \ell_F=
  \begin{pmatrix}
    \sum_i\pd{F^1}{u_i^1}D_x^i&\dots&\sum_i\pd{F^1}{u_i^m}D_x^i\\
    \hdotsfor{3}\\
    \sum_i\pd{F^m}{u_i^1}D_x^i&\dots&\sum_i\pd{F^m}{u_i^m}D_x^i
  \end{pmatrix}.
\end{equation}
Operator~\eqref{eq:bouss-new:6} is called the \emph{linearization
  operator} for the equation~$\mathcal{E}$. The set of symmetries is a
Lie algebra denoted by~$\sym(\mathcal{E})$.

A \emph{conservation law} for equation~\eqref{eq:bouss-new:2} is a
horizontal $1$-form
\begin{equation*}
  \omega=X\,dx+T\,dt
\end{equation*}
closed with respect to the horizontal de~Rham differential
\begin{equation*}
  d_h=dx\wedge D_x+dt\wedge D_t,
\end{equation*}
i.e., such that
\begin{equation*}
  D_x(T)=D_t(X),
\end{equation*}
where~$T=T(x,t,u,u_1,\dots)$, $X=X(x,t,u,u_1,\dots)$. A conservation
law is \emph{trivial} if it is of the from~$\omega=d_hf$, i.e.,
\begin{equation*}
  X=D_x(f),\quad T=D_t(f),\qquad f=f(x,t,u,u_1,\dots).
\end{equation*}
The space of equivalence classes of conservation laws modulo trivial
ones coincides with the $1$st horizontal de~Rham cohomology group
for~$\mathcal{E}$ and is denoted by~$H_h^1(\mathcal{E})$.

\begin{remark}
  If the number of the space variables~$x$ equals~$n$, then this space
  coincides with~$H_h^n(\mathcal{E})$.
\end{remark}

To any conservation law~$\omega=X\,dx+T\,dt$ there corresponds its
\emph{generating function}
\begin{equation}
  \label{eq:bouss-new:8}
  \psi_\omega=\delta X=\Big(\fd{X}{u^1},\dots,\fd{X}{u^m}\Big),
\end{equation}
where~$\delta$ denotes the \emph{Euler operator} and
\begin{equation}
  \label{eq:bouss-new:9}
  \fd{}{u^j}=\sum_{i\ge0}(-1)^iD_x^i\circ\pd{}{u_i^j}
\end{equation}
is the \emph{variational derivative} with respect to~$u^j$. Any
generating function~\eqref{eq:bouss-new:8} satisfies the equation
\begin{equation}
  \label{eq:bouss-new:10}
  \ell_{\mathcal{E}}^*(\psi_\omega)=0,
\end{equation}
where
\begin{multline}
  \label{eq:bouss-new:11}
  \ell_{\mathcal{E}}^*=-D_t+\ell_F^*\\=-D_t+
  \begin{pmatrix}
    \sum_i(-1)^iD_x^i\circ\pd{F^1}{u_i^1}&\dots&
    \sum_i(-1)^iD_x^i\circ\pd{F^m}{u_i^1}\\
    \hdotsfor{3}\\
    \sum_i(-1)^iD_x^i\circ\pd{F^1}{u_i^m}&\dots&
    \sum_i(-1)^iD_x^i\circ\pd{F^m}{u_i^m}
  \end{pmatrix}
\end{multline}
is the operator adjoint to~$\ell_{\mathcal{E}}$.

Solutions of equation~\eqref{eq:bouss-new:10} are called
\emph{cosymmetries} and in general not all of them are generating
functions of conservation laws. The space of cosymmetries will be
denoted by~$\sym^*(\mathcal{E})$. The Euler operator determines the
embedding
\begin{equation}
  \label{eq:bouss-new:12}
  \delta\colon H_h^1(\mathcal{E})\to\sym^*(\mathcal{E}).
\end{equation}

Symmetries and cosymmetries may be understood as vector fields and
differential $1$-forms, respectively, on the equation~$\mathcal{E}$
and there is a natural pairing between them:
if~$\phi=(\phi^1,\dots,\phi^m)\in\sym(\mathcal{E})$
and~$\psi=(\psi^1,\dots,\psi^m)\in\sym^*(\mathcal{E})$, we set
\begin{equation}
  \label{eq:bouss-new:13}
  \langle\psi,\phi\rangle=\psi^1\phi^1+\dots+\psi^m\phi^m.
\end{equation}
At first glance, the right-hand side of~\eqref{eq:bouss-new:13} looks
like a function, but the ``physical meaning''
of~$\langle\psi,\phi\rangle$ is quite different. Namely,
applying~$D_t$ to~$\langle\psi,\phi\rangle$ we have
\begin{equation*}
  D_t\langle\psi,\phi\rangle=\langle D_t(\psi),\phi\rangle+
  \langle\psi,D_t(\phi)\rangle=
  -\langle\ell_F^*(\psi),\phi\rangle+
  \langle\psi,\ell_F(\phi)
  \rangle
\end{equation*}
and, consequently,
\begin{equation*}
  D_t\langle\psi,\phi\rangle=D_x(T_{\psi,\phi})
\end{equation*}
for some~$T_{\psi,\phi}$. Though the conservation law
\begin{equation}
  \label{eq:bouss-new:18}
  \langle\psi,\phi\rangle\,dx+T_{\psi,\phi}\,dt
\end{equation}
is not defined uniquely, its cohomology class depends on~$\phi$
and~$\psi$ only and we obtain
\begin{equation*}
  \langle\cdot\,,\cdot\rangle\colon
  \sym^*(\mathcal{E})\times\sym(\mathcal{E})\to H_h^1(\mathcal{E}).
\end{equation*}
Note that for any~$\omega\in H_h^1(\mathcal{E})$ one has
\begin{equation*}
  \langle\delta\omega,\phi\rangle=\re_\phi(\omega).
\end{equation*}

\subsection{Recursion operators, Hamiltonian and symplectic
  structures}
\label{sec:recurs-oper-hamilt}

In this subsection we shall discuss a local theory and shall explain
how nonlocal components are incorporated in all construction in
Subsection~\ref{sec:nonlocal-theory}. All operators considered below
are matrix operators in total derivatives of the form
\begin{equation}
  \label{eq:bouss-new:14}
  \Delta=
  \begin{pmatrix}
    \sum_{i,s}a_{jl}^{is}D_x^iD_t^s
  \end{pmatrix}.
\end{equation}
We call such operators \emph{$\mathcal{C}$-differential operators}. In
particular, we shall deal with operators
\begin{equation}
  \label{eq:bouss-new:15}
  \Delta=
  \begin{pmatrix}
    \sum_i a_{jl}^i D_x^i
  \end{pmatrix}
\end{equation}
in~$D_x$ only.

In the case of equation~\eqref{eq:bouss-new:2} we may consider the
operator~$\ell_{\mathcal{E}}$ to act from the space~$\kappa$ of vector
functions~$\phi=(\phi^1,\dots,\phi^m)$ to the same space.
Then~$\ell_{\mathcal{E}}^*$ acts from~$\kappa^*$ to~$\kappa^*$,
where~$\kappa^*$ is the dual space.

\begin{remark}
  A coordinate-free description of actions of~$\ell_{\mathcal{E}}$
  and~$\ell_{\mathcal{E}}^*$ may be found
  in~\cite{KrasilshchikVinogradov:SCLDEqMP}.
\end{remark}

From Subsection~\ref{subsec:symm-cons-laws} we have
\begin{equation*}
  \sym(\mathcal{E})=\ker\ell_{\mathcal{E}}\subset\kappa,\qquad
  \sym^*(\mathcal{E})=\ker\ell_{\mathcal{E}}^*\subset\kappa^*
\end{equation*}
and we are looking for $\mathcal{C}$-differential operators of the
form~\eqref{eq:bouss-new:15} acting as follows
\begin{align}
  \label{eq:bouss-new:16}
  &\mathcal{R}\colon\kappa\to\kappa,\\
  \label{eq:bouss-new:19}
  &\mathcal{H}\colon\kappa^*\to\kappa,\\
  \label{eq:bouss-new:20}
  &\mathcal{S}\colon\kappa\to\kappa^*,\\
  \label{eq:bouss-new:21}
  &\mathcal{R}^*\colon\kappa^*\to\kappa^*,
\end{align}
and such that
\begin{align}
  \label{eq:bouss-new:17}
  &\mathcal{R}(\sym(\mathcal{E}))\subset\sym(\mathcal{E}),\\
  \label{eq:bouss-new:22}
  &\mathcal{H}(\sym^*(\mathcal{E}))\subset\sym(\mathcal{E}),\\
  \label{eq:bouss-new:23}
  &\mathcal{S}(\sym(\mathcal{E}))\subset\sym^*(\mathcal{E}),\\
  \label{eq:bouss-new:24}
  &\mathcal{R}^*(\sym^*(\mathcal{E}))\subset\sym^*(\mathcal{E}).
\end{align}
To find such operators makes the first step in constructing the
structures we are interested in. Namely,
\begin{itemize}
\item operators~$\mathcal{R}$ are recursion operators for symmetries;
\item operators~$\mathcal{H}$ satisfying conditions described in
  Remark~\ref{rem:recurs-oper-hamilt-1} below are Hamiltonian
  structures on the equation~$\mathcal{E}$;
\item operators~$\mathcal{S}$ satisfying conditions described in
  Remark~\ref{rem:recurs-oper-hamilt-2} below are symplectic
  structures on the equation~$\mathcal{E}$;
\item operators~$\mathcal{R}^*$.
\end{itemize}

\begin{remark}[Hamiltonianity conditions]
  \label{rem:recurs-oper-hamilt-1}
  Let~$\mathcal{H}$ be an operator of
  type~\eqref{eq:bouss-new:19}. Then for any two conservation
  laws~$\omega_1$ and~$\omega_2$ of equation~\eqref{eq:bouss-new:2}
  (or their equivalence classes in~$H_h^1(\mathcal{E})$) one can
  define the bracket
  \begin{equation}
    \label{eq:bouss-new:25}
    \{\omega_1,\omega_2\}_{\mathcal{H}}=
    \langle\delta\omega_1,\mathcal{H}(\delta\omega_2)\rangle=
    \re_{\mathcal{H}(\delta\omega_2)}(\delta\omega_1).
  \end{equation}
  $\mathcal{H}$ is a \emph{Hamiltonian structure} if this bracket is
  skew-symmetric and satisfies the Jacobi identity. The fist condition
  means that~$\mathcal{H}$ is a skew-adjoint operator
  ($\mathcal{H}=-\mathcal{H}^*$) and thus may be understood as a
  \emph{bivector} on~$\mathcal{E}$, while the second one amounts to
  \begin{equation}
    \label{eq:bouss-new:26}
    \ldb\mathcal{H},\mathcal{H}\rdb=0,
  \end{equation}
  where~$\ldb\cdot,\cdot\rdb$ is the \emph{variational Schouten
    bracket},~\cite{IgoninVerbovetskyVitolo:FLVDOp}. Recall also that
  two Hamiltonian structures are \emph{compatible} (or constitute a
  \emph{Hamiltonian pair}) if and only if
  \begin{equation}
    \label{eq:bouss-new:27}
    \ldb\mathcal{H}_1,\mathcal{H}_2\rdb=0.
  \end{equation}
  Efficient coordinate formulas to check
  conditions~\eqref{eq:bouss-new:26} and~\eqref{eq:bouss-new:27}
  will be given in Subsection~\ref{sec:gener-outl-comp}.
\end{remark}

\begin{remark}[symplectic conditions]
  \label{rem:recurs-oper-hamilt-2}
  Let~$\mathcal{S}$ be an operator of
  type~\eqref{eq:bouss-new:20}. Then it can be understood as a map
  \begin{equation}
    \label{eq:bouss-new:28}
    \mathcal{S}\colon\sym(\mathcal{E})\times\sym(\mathcal{E})
    \to H_h^1(\mathcal{E})
  \end{equation}
  by setting
  \begin{equation}
    \label{eq:bouss-new:29}
    \mathcal{S}(\phi_1,\phi_2)=
    \langle\mathcal{S}(\phi_1),\phi_2\rangle.
  \end{equation}
  We say that~$\mathcal{S}$ is a \emph{two-form} on~$\mathcal{E}$ if
  \begin{equation}
    \label{eq:bouss-new:30}
    \mathcal{S}(\phi_1,\phi_2)=-\mathcal{S}(\phi_2,\phi_1)
  \end{equation}
  for all~$\phi_1$, $\phi_2\in\sym(\mathcal{E})$ (this is equivalent
  to~$\mathcal{S}=-\mathcal{S}^*$). For such a form one can define its
  differential
  \begin{equation*}
    \delta\mathcal{S}\colon\sym(\mathcal{E})\times
    \sym(\mathcal{E})\times\sym(\mathcal{E})
    \to H_h^1(\mathcal{E})
  \end{equation*}
  by
  \begin{multline}
    \label{eq:bouss-new:31}
    \delta\mathcal{S}(\phi_1,\phi_2,\phi_3)=
    \re_{\phi_1}\mathcal{S}(\phi_2,\phi_3)-
    \re_{\phi_2}\mathcal{S}(\phi_1,\phi_3)-
    \re_{\phi_3}\mathcal{S}(\phi_1,\phi_2)\\
    -\mathcal{S}(\{\phi_1,\phi_2\},\phi_3)
    +\mathcal{S}(\{\phi_1,\phi_3\},\phi_2)
    -\mathcal{S}(\{\phi_2,\phi_3\},\phi_3),
  \end{multline}
  where the bracket~$\{\phi,\phi'\}$ is uniquely defined by the
  formula
  \begin{equation*}
    \re_{\{\phi,\phi'\}}=[\re_\phi,\re_{\phi'}].
  \end{equation*}
  We say that~$\mathcal{S}$ is a \emph{symplectic structure} if the
  corresponding two-form is closed with respect to~$\delta$, i.e.,
  \begin{equation}
    \label{eq:bouss-new:32}
    \delta\mathcal{S}=0.
  \end{equation}
  Again, computational formulas to check
  conditions~\eqref{eq:bouss-new:30} and~\eqref{eq:bouss-new:32}
  will be given in Subsection~\ref{sec:gener-outl-comp}.
\end{remark}

\begin{remark}[Nijenhuis conditions]
  \label{rem:recurs-oper-hamilt-3}
  Any operator~\eqref{eq:bouss-new:16}
  satisfying~\eqref{eq:bouss-new:17} can be considered as a recursion
  on symmetries, i.e., it takes a symmetry of the
  equation~$\mathcal{E}$ to a symmetry. Similarly, any
  operator~\eqref{eq:bouss-new:24} is a recursion operator for
  cosymmetries. Thus, applying~$\mathcal{R}$ to a given
  symmetry~$\phi$ one obtains the whole family
  \begin{equation}
    \label{eq:bouss-new:34}
    \phi_0=\phi,\ \phi_1=\mathcal{R}\phi,\dots,
    \phi_i=\mathcal{R}^i\phi,\dots
  \end{equation}
  Commutativity of such families closely relates to integrability
  of~$\mathcal{E}$.

  For any recursion
  operator~$\mathcal{R}\colon\sym(\mathcal{E})\to\sym(\mathcal{E})$
  its \emph{Nijenhuis torsion}
  \begin{equation}
    \label{eq:bouss-new:33}
    N_{\mathcal{R}}\colon\sym(\mathcal{E})\times\sym(\mathcal{E})
    \to\sym(\mathcal{E})
  \end{equation}
  is defined by
  \begin{equation*}
    N_{\mathcal{R}}(\phi_1,\phi_2)=
    \{\mathcal{R}\phi_1,\mathcal{R}\phi_1\}-
    (\mathcal{R}(\{\mathcal{R}\phi_1,\phi_2\})-
    \{\phi_1,\mathcal{R}\phi_2\}+\mathcal{R}(\{\phi_1,\phi_2\})),
  \end{equation*}
  while for a symmetry~$\phi$ its action on~$\mathcal{R}$ is defined
  by
  \begin{equation*}
    (\re_\phi\mathcal{R})(\phi')=
    \{\phi,\mathcal{R}\phi'\}-\mathcal{R}\{\phi,\phi'\}.
  \end{equation*}
  Both operations can be expressed in terms of the \emph{Nijenhuis
    bracket} and as it was shown in~\cite{KKLS} the conditions
  \begin{equation}
    \label{eq:bouss-new:35}
    N_{\mathcal{R}}=0,\qquad\re_\phi\mathcal{R}=0
  \end{equation}
  imply commutativity of family~\eqref{eq:bouss-new:34}.
\end{remark}

\subsection{$\Delta$-coverings}
\label{sec:delta-coverings}

To construct operators possessing
properties~\eqref{eq:bouss-new:17}--\eqref{eq:bouss-new:24}, we
solve the following operator equations
\begin{align}
  \label{eq:bouss-new:36}
  &\ell_{\mathcal{E}}\circ\mathcal{R}=
  A_{\mathcal{R}}\circ\ell_{\mathcal{E}}\\
  \label{eq:bouss-new:37}
  &\ell_{\mathcal{E}}\circ\mathcal{H}=
  A_{\mathcal{H}}\circ\ell_{\mathcal{E}}^*\\
  \label{eq:bouss-new:38}
  &\ell_{\mathcal{E}}^*\circ\mathcal{S}=
  A_{\mathcal{S}}\circ\ell_{\mathcal{E}}\\
  \label{eq:bouss-new:39}
  &\ell_{\mathcal{E}}^*\circ\mathcal{R}^*=
  A_{\mathcal{R}^*}\circ\ell_{\mathcal{E}}^*
\end{align}
with respect to~$\mathcal{R}$, $\mathcal{H}$, $\mathcal{S}$,
and~$\mathcal{R}^*$ for some $\mathcal{C}$-differential
operators~$A_{\mathcal{R}}$, $A_{\mathcal{H}}$, $A_{\mathcal{S}}$,
and~$A_{\mathcal{R}^*}$.

\begin{remark}
  It is easily shown that
  \begin{equation*}
    A_{\mathcal{R}}=\mathcal{R},\ A_{\mathcal{H}}=-\mathcal{H},\
    A_{\mathcal{S}}=-\mathcal{S},\ A_{\mathcal{R}^*}=\mathcal{R}^*
  \end{equation*}
  and for any solution~$\mathcal{R}$ of
  equation~\eqref{eq:bouss-new:36} its adjoint~$\mathcal{R}^*$ is a
  solution of~\eqref{eq:bouss-new:39} and vice versa, while for any
  solution~$\mathcal{H}$ of equation~\eqref{eq:bouss-new:37} its
  inverse~$\mathcal{H}^{-1}$ (if it makes sense) is a solution
  of~\eqref{eq:bouss-new:38} and vice versa. In addition, for any
  solution~$\mathcal{H}$ its adjoint~$\mathcal{H}^*$ is a solution as
  well and the same for~$\mathcal{S}$.
\end{remark}

All four problems~\eqref{eq:bouss-new:36}--\eqref{eq:bouss-new:39}
can be formulated in the following general form. Let~$P$, $Q$, $P'$,
$Q'$ be some spaces of vector functions on the equation~$\mathcal{E}$
and~$\Delta\colon P\to P'$, $\nabla\colon Q\to Q'$ be
$\mathcal{C}$-differential operators. How to find all
$\mathcal{C}$-differential operators~$\mathcal{X}\colon P\to Q$ such
that the diagram
\begin{equation*}
  \begin{CD}
    P@>\mathcal{X}>>Q\\
    @V{\Delta}VV@VV{\nabla}V\\
    P'@>{A_{\mathcal{X}}}>>Q'
  \end{CD}
\end{equation*}
is commutative for some $\mathcal{C}$-differential
operator~$A_{\mathcal{X}}$? For simplicity, we shall assume
that~$\mathcal{X}$ is an operator in~$D_x$ only, i.e., of the
form~\eqref{eq:bouss-new:15}.

\begin{remark}
  This problem has an obvious set of trivial solutions. Namely,
  any~$\mathcal{X}$ of the form~$\mathcal{X}=\mathcal{X}'\circ\Delta$.
  We shall look for equivalence classes of solutions modulo trivial
  ones.
\end{remark}

To this end, we use the following construction. Assume that the
operator~$\Delta$ acts on functions~$v=(v^1,\dots,v^r)$ and consider
the system of equations
\begin{equation}
  \label{eq:bouss-new:40}
  \begin{cases}
    u_t=F(x,t,u,u_1,\dots,u_k),\\
    \Delta(v)=0
  \end{cases}
\end{equation}
(recall that the coefficient of~$\Delta$ may depend on the
function~$u$ and its derivatives). We say that
system~\eqref{eq:bouss-new:40} is the \emph{$\Delta$-covering} (or
the \emph{$\Delta$-extension}) of the equation~$\mathcal{E}$.

\begin{example}[the $\ell_{\mathcal{E}}$-covering]
  Let us take the operator~$\ell_{\mathcal{E}}=D_t-\ell_F$
  for~$\Delta$. Then the corresponding extension is called the
  \emph{$\ell_{\mathcal{E}}$-covering} of~$\mathcal{E}$. For example,
  if~$\mathcal{E}$ is the Burgers equation
  \begin{equation*}
    u_t=u_{xx}+uu_x,
  \end{equation*}
  then~$\ell_{\mathcal{E}}=D_t-D_x^2-uD_x-u_x$ and the
  $\ell_{\mathcal{E}}$ covering is of the form
  \begin{equation*}
    \begin{cases}
      u_t=u_{xx}+uu_x,\\
      v_t=v_{xx}+uv_x+vu_x.
    \end{cases}
  \end{equation*}
\end{example}

\begin{example}[the $\ell_{\mathcal{E}}^*$-covering]
  If we take the operator~$\ell_{\mathcal{E}}^*=-D_t-\ell_F^*$
  for~$\Delta$, then this extension is called the
  \emph{$\ell_{\mathcal{E}}^*$-covering}. For the Burgers equation,
  $\ell_{\mathcal{E}}^*=-D_t-D_x^2+D_x\circ u-u_x=-D_t-D_x^2+uD_x$ and
  thus
  \begin{equation*}
    \begin{cases}
      u_t=u_{xx}+uu_x,\\
      v_t=-v_{xx}+uv_x
    \end{cases}
  \end{equation*}
  is the $\ell_{\mathcal{E}}^*$-covering.
\end{example}

Denote by~$\tilde{D}_x$ and~$\tilde{D}_t$ the total derivatives in the
space of the $\Delta$-covering. Then, since~$\nabla$ is a
$\mathcal{C}$-differential operator, we can consider the
operator~$\tilde{\nabla}$ which is obtained from~$\nabla$ by
changing~$D_x$ and~$D_t$ to~$\tilde{D}_x$ and~$\tilde{D}_t$,
respectively. Consider now the linear differential equation
\begin{equation}
  \label{eq:bouss-new:41}
  \tilde{\nabla}(a)=0,
\end{equation}
where~$a=(a^1,\dots,a^s)$ is a vector function linear in the
variables~$v$, $v_1,\dots,v_i,\dots$, i.e.,
\begin{equation}
  \label{eq:bouss-new:42}
  a^j=\sum_{i,l}a_{il}^jv_i^l,
\end{equation}
$a_{il}^j$ being functions on~$\mathcal{E}$.

\begin{theorem}
  Equivalence classes of solutions of the equation
  \begin{equation*}
    \nabla\circ\mathcal{X}=A_{\mathcal{X}}\circ\Delta
  \end{equation*}
  modulo trivial solutions are in one-to-correspondence with the
  solutions of equation~\eqref{eq:bouss-new:41} of the
  form~\eqref{eq:bouss-new:42}. The operator~$\mathcal{X}$
  corresponding to solution~\eqref{eq:bouss-new:42} is of the form
  \begin{equation}
    \label{eq:bouss-new:43}
    \mathcal{X}=
    \begin{pmatrix}
      \sum_ia_{il}^jD_x^i
    \end{pmatrix}.
  \end{equation}
\end{theorem}

\subsection{Nonlocal theory}
\label{sec:nonlocal-theory}

$\Delta$-coverings are a particular case of the general geometric
construction that allows to introduce nonlocal objects related to the
initial differential equation
(see~\cite{KrasilshchikVinogradov:NTGDEqSCLBT}). Here we shall present
all needed constructions in the coordinate form.

Let~$\mathcal{E}$ be an equation of the form~\eqref{eq:bouss-new:2}
and~$w=(w^1,\dots,w^r)$ be a new unknown vector function in~$x$
and~$t$ (the case of infinite number of~$w$'s is included). A
system~$\tilde{\mathcal{E}}$
\begin{equation}
  \label{eq:bouss-new:44}
  \begin{cases}
    w_x^j=X^j(x,t,w,u,u_1,\dots,u_p),\\
    w_t^j=T^j(x,t,w,u,u_1,\dots,u_p),
  \end{cases}
\end{equation}
$j=1,\dots,r$, is called a \emph{covering} over~$\mathcal{E}$ if its
compatibility is provided by~$\mathcal{E}$. The number~$r$ of new
variables is called the \emph{dimension} of the covering while~$w$'s
are said to be \emph{nonlocal variables} in this covering.

\begin{example}
  The system
  \begin{equation*}
    w_x=u,\qquad w_t=u_{xx}+\frac{1}{2}u^2
  \end{equation*}
  is a covering over the KdV equation.
\end{example}

Total derivatives on~$\tilde{\mathcal{E}}$ are denote~$\tilde{D}_x$
and~$\tilde{D}_t$ and they are of the form
\begin{equation*}
  \tilde{D}_x=D_x+\sum_{j=1}^mX^j\pd{}{w^j},\qquad
  \tilde{D}_t=D_t+\sum_{j=1}^mT^j\pd{}{w^j}.
\end{equation*}
Thus, if~$\Delta$ is a $\mathcal{C}$-differential operator
on~$\mathcal{E}$, we can define the operator~$\tilde{\Delta}$
on~$\tilde{\mathcal{E}}$ substituting~$D_x$ by~$\tilde{D}_x$ and~$D_t$
by~$\tilde{D}_t$.

Compatibility conditions for~\eqref{eq:bouss-new:44} read
\begin{equation*}
  \tilde{D}_tX^j=\tilde{D}_xT^j\bmod\tilde{\mathcal{E}}.
\end{equation*}
If~$X^j$ and~$T^j$ do not depend on~$w$, then these conditions mean
that the forms
\begin{equation*}
  \omega^1=X^1\,dx+T^1\,dt,\dots,\omega^j=X^j\,dx+T^j\,dt,\dots
\end{equation*}
are conservation laws for the equation~$\mathcal{E}$. Conversely, to
any system of conservation laws we can put into correspondence a
covering.

The system~$\tilde{\mathcal{E}}$ is a differential equation itself and
thus we can consider its symmetries, cosymmetries, conservation laws,
etc. These objects will be referred to as \emph{nonlocal} with respect
to the initial equation~$\mathcal{E}$. In particular, nonlocal
symmetries are determined by the equation
\begin{equation*}
  \ell_{\tilde{\mathcal{E}}}(\tilde{\phi})=0,
\end{equation*}
while nonlocal cosymmetries are defined by
\begin{equation*}
  \ell_{\tilde{\mathcal{E}}}^*(\tilde{\psi})=0.
\end{equation*}
On the other hand, since~$\ell_{\mathcal{E}}$ is a
$\mathcal{C}$-differential operator, the
operator~$\tilde{\ell}_{\mathcal{E}}$ is well defined. In general,
$\ell_{\tilde{\mathcal{E}}}\neq\tilde{\ell}_{\mathcal{E}}$ and the
equations
\begin{equation*}
  \tilde{\ell}_{\mathcal{E}}(\tilde{\phi})=0
\end{equation*}
and
\begin{equation*}
  \tilde{\ell}_{\mathcal{E}}^*(\tilde{\psi})=0
\end{equation*}
can be considered. Their solutions are called \emph{shadows} of
symmetries and cosymmetries, respectively.

\begin{remark}
  $\Delta$-coverings are (usually, infinite-dimensional) coverings in
  the sense of this subsection. Consider, for example, the
  $\ell_{\mathcal{E}}$-covering which can be presented in the form
  \begin{equation*}
    u_t=F(x,t,u,u_1,\dots,u_k),\qquad
    v_t=\ell_F(v)
  \end{equation*}
  and set
  \begin{equation*}
    \omega^0=v,\ \omega^1=v_x,\ \omega^2=v_{xx},\dots
  \end{equation*}
  Then the $\ell_{\mathcal{E}}$-covering is equivalent to the system
  \begin{equation*}
    \begin{cases}
      \omega_x^j=\omega^{j+1},\\
      \omega_t^j=\tilde{D}_x^j(\ell_F(\omega^0)),
    \end{cases}
  \end{equation*}
  $j=0,1,2,\dots$ In a similar way, the
  $\ell_{\mathcal{E}}^*$-covering is equivalent to
  \begin{equation*}
    \begin{cases}
      \rho_x^j=\rho^{j+1},\\
      \rho_t^j=-\tilde{D}_x^j(\ell_F^*(\rho^0)).
    \end{cases}
  \end{equation*}
  From now on, the notation~$\omega$ and~$\rho$ is used for nonlocal
  variables in the~$\ell_{\mathcal{E}}$- and
  $\ell_{\mathcal{E}}^*$-coverings, respectively.
\end{remark}

Comparing the results of this and the previous subsections, we can
formulate the statement fundamental for all subsequent computations.
\begin{theorem}
  \label{thm:nonlocal-theory-1}
  Let~$\mathcal{E}$ be an equation of the
  form~\eqref{eq:bouss-new:2}. Then equivalence classes of
  solutions~$\mathcal{R}$\textup{,} $\mathcal{H}$\textup{,}
  $\mathcal{S}$\textup{,} $\mathcal{R}^*$ of
  equations~\eqref{eq:bouss-new:36}\textup{,}
  \eqref{eq:bouss-new:37}\textup{,}
  \eqref{eq:bouss-new:38}\textup{,} \eqref{eq:bouss-new:39} are in
  one-to-one correspondence with
  \begin{itemize}
  \item shadows of symmetries in the $\ell_{\mathcal{E}}$-covering
    for~\eqref{eq:bouss-new:36}\textup{,}
  \item shadows of symmetries in the $\ell_{\mathcal{E}}^*$-covering
    for~\eqref{eq:bouss-new:37}\textup{,}
  \item shadows of cosymmetries in the $\ell_{\mathcal{E}}$-covering
    for~\eqref{eq:bouss-new:38}\textup{,}
  \item shadows of cosymmetries in the $\ell_{\mathcal{E}}^*$-covering
    for~\eqref{eq:bouss-new:39}.
  \end{itemize}
  All shadows are taken to be linear with respect to nonlocal
  variables in the corresponding covering.
\end{theorem}

Since~$\tilde{\mathcal{E}}$ is an equation, we can consider its
coverings. In this way, coverings over coverings, etc., arise.  Our
next remark is concerned with special coverings
over~$\ell_{\mathcal{E}}$- and $\ell_{\mathcal{E}}^*$-coverings and
with interpretation of the corresponding nonlocal variables.

\begin{remark}[nonlocal vectors and covectors]
  \label{rem:nonlocal-vect-covect}
  Consider an equation~$\mathcal{E}$ and its
  $\ell_{\mathcal{E}}^*$-covering. Let~$\phi$ be a symmetry
  of~$\mathcal{E}$. Then, since~$\phi$ satisfies the
  equation~$D_t\phi=\ell_F\phi$ and~$\rho$ is defined
  by~$D_t\rho=-\ell_F^*\rho$, we have
  \begin{equation*}
    D_t(\rho\phi)=D_x(T_\phi)
  \end{equation*}
  for some function~$T_\phi$ on the $\ell_{\mathcal{E}}^*$-covering
  (see the end of Subsection~\ref{subsec:symm-cons-laws}).
  Consequently,
  \begin{equation*}
    \rho\phi\,dx+T_\phi\,dt
  \end{equation*}
  is a conservation law on the $\ell_{\mathcal{E}}^*$-covering. Denote
  by~$\rho^\phi$ the nonlocal variable in the corresponding
  covering,~i.e.,
  \begin{equation}
    \label{eq:bouss-new:45}
    \rho_x^\phi=\rho\phi,\qquad\rho_t^\phi=T_\phi.
  \end{equation}
  We call~$\rho^\phi$ the \emph{nonlocal vector} corresponding
  to~$\phi$.

  In a similar way, any cosymmetry~$\psi$ of~$\mathcal{E}$ determines
  the conservation law
  \begin{equation*}
    \psi\omega\,dx+T_\psi\,dt
  \end{equation*}
  on the $\ell_{\mathcal{E}}$-covering. The nonlocal
  variable~$\omega^\psi$ defined by
  \begin{equation}
    \label{eq:bouss-new:46}
    \omega_x^\psi=\psi\omega,\qquad\omega_t^\psi=T_\psi
  \end{equation}
  is called the \emph{nonlocal covector} (or \emph{nonlocal form})
  corresponding to~$\psi$.

  Take a set of symmetries $\phi^1,\dots,\phi^r$ and the covering
  corresponding to this set by~\eqref{eq:bouss-new:45}. Then we can
  consider shadows of symmetries and cosymmetries in this covering
  linear with respect to the variables~$\rho$ and~$\rho^\phi$,~i.e.,
  shadows of the form~$a=(a^1,\dots,a^m)$, where
  \begin{equation}
    \label{eq:bouss-new:47}
    a^j=\sum_{l=1}^m\sum_{i\ge0}a_i^{jl}\rho_i^l
    +\sum_{s=1}^r a_s^j\rho^{\phi_s}.
  \end{equation}
  
  \begin{proposition}
    \label{prop:nonlocal-theory-1}
    The operators corresponding to functions~\eqref{eq:bouss-new:47}
    are of the form
    \begin{equation}
      \label{eq:bouss-new:48}
      \Delta=
      \begin{pmatrix}
        \sum_{i\ge0}a_i^{jl}D_x^i+
        \sum_{s=1}^r a_s^jD_x^{-1}\circ\phi_s^l
      \end{pmatrix},\qquad
      j,l=1,\dots,m,
    \end{equation}
    and they enjoy properties~\eqref{eq:bouss-new:37}
    and~\eqref{eq:bouss-new:39} for shadows of symmetries and
    cosymmetries\textup{,} respectively. In addition\textup{,}
    if~\eqref{eq:bouss-new:47} is a shadow of symmetry\textup{,} then
    the vector functions
    \begin{equation*}
      a_s=(a_s^1,\dots,a_s^m),\qquad s=1,\dots,r,
    \end{equation*}
    are symmetries of equation~$\mathcal{E}$\textup{,} and
    if~\eqref{eq:bouss-new:47} is a shadow of cosymmetry\textup{,}
    then these functions are cosymmetries.
  \end{proposition}

  In the same way, one can consider a set~$\psi_1,\dots,\psi_r$ of
  cosymmetries and the covering over the $\ell_{\mathcal{E}}$-covering
  given by~\eqref{eq:bouss-new:46}. Let us take a
  shadow~$b=(b^1,\dots,b^m)$,
  \begin{equation}
    \label{eq:bouss-new:49}
    b^j=\sum_{l=1}^m\sum_{i\ge0}b_i^{jl}\omega_i^l+
    \sum_{s=1}^rb_s^j\omega^{\psi_s}
  \end{equation}
in this covering.
  
  \begin{proposition}
    \label{prop:nonlocal-theory-2}
    The operators corresponding to functions~\eqref{eq:bouss-new:49}
    are of the form
    \begin{equation}
      \label{eq:bouss-new:50}
      \Delta=
      \begin{pmatrix}
        \sum_{i\ge0}b_i^{jl}D_x^i+
        \sum_{s=1}^r b_s^jD_x^{-1}\circ\psi_s^l
      \end{pmatrix},\qquad
      j,l=1,\dots,m,
    \end{equation}
    and they enjoy properties~\eqref{eq:bouss-new:36}
    and~\eqref{eq:bouss-new:38} for shadows of symmetries and
    cosymmetries\textup{,} respectively. In addition\textup{,}
    if~\eqref{eq:bouss-new:49} is a shadow of symmetry\textup{,} then
    the vector functions
    \begin{equation*}
      b_s=(b_s^1,\dots,b_s^m),\qquad s=1,\dots,r,
    \end{equation*}
    are symmetries of~$\mathcal{E}$\textup{,} and
    if~\eqref{eq:bouss-new:49} is a shadow of cosymmetry\textup{,}
    then these functions are cosymmetries.
  \end{proposition}
  
  Note that all operators~\eqref{eq:bouss-new:48}
  and~\eqref{eq:bouss-new:50} are obtained in the \emph{weakly
    nonlocal form} (cf.~\cite{Rubtsov-Orlov}).
\end{remark}

\begin{remark}
  \label{rem:nonlocal-theory-notation}
  In the sequel, we shall use a short notation
  \begin{equation*}
    \langle a_1,\dots,a_r\mid D_x^{-1}\mid\phi_1,\dots,\phi_r\rangle
    =\sum_{s=1}^r a_s^jD_x^{-1}\circ\phi_s^l
  \end{equation*}
  and
  \begin{equation*}
    \langle b_1,\dots,b_r\mid D_x^{-1}\mid\psi_1,\dots,\psi_r\rangle
    =\sum_{s=1}^r b_s^jD_x^{-1}\circ\psi_s^l
  \end{equation*}
  for the nonlocal part of operators~\eqref{eq:bouss-new:48}
  and~\eqref{eq:bouss-new:50}, resp.
\end{remark}

\subsection{General outline of the computational scheme}
\label{sec:gener-outl-comp}

The computations described in Sections~\ref{sec:prep-comp}
and~\ref{sec:main-results} are divided into two parts. The first one
plays an auxiliary role and consists of the following steps:
\begin{itemize}
\item Construction of a number of conservation laws and the
  corresponding nonlocal variables for the initial equation.
\item Construction of local and nonlocal symmetries that will serve as
  seed symmetries for infinite families and to be used to construct
  necessary nonlocal vectors.
\item Construction of local and nonlocal cosymmetries that will serve as
  seed cosymmetries for infinite families and to be used to construct
  necessary nonlocal covectors.
\item Construction of nonlocal vectors and covectors.
\end{itemize}

In the second part we reveal the main structures associated to the
dB-equation. Namely,
\begin{itemize}
\item We solve the equation
  \begin{equation*}
    \tilde{\ell}_{\mathcal{E}}(b)=0
  \end{equation*}
  in the $\ell_{\mathcal{E}}$-covering to construct \emph{recursion
    operators for symmetries} of the form~\eqref{eq:bouss-new:50}.
\item We solve the equation
  \begin{equation*}
    \tilde{\ell}_{\mathcal{E}}^*(a)=0
  \end{equation*}
  in the $\ell_{\mathcal{E}}^*$-covering to construct \emph{recursion
    operators for cosymmetries} of the form~\eqref{eq:bouss-new:48}.
\item We solve the equation
  \begin{equation}
    \label{eq:bouss-new:51}
    \tilde{\ell}_{\mathcal{E}}(a)=0
  \end{equation}
  in the $\ell_{\mathcal{E}}^*$-covering to find operators of the
  form~\eqref{eq:bouss-new:48} that take cosymmetries to
  symmetries. \emph{Hamiltonian structures} on~$\mathcal{E}$ are
  distinguished by Theorem~\ref{thm:gener-outl-comp-1} below.
\item We solve the equation
  \begin{equation}
    \label{eq:bouss-new:52}
    \tilde{\ell}_{\mathcal{E}}^*(b)=0
  \end{equation}
  in the $\ell_{\mathcal{E}}$-covering to find operators of the
  form~\eqref{eq:bouss-new:50} that take symmetries to cosymmetries.
  Symplectic structures on~$\mathcal{E}$ are distinguished by
  Theorem~\ref{thm:gener-outl-comp-2}.
\end{itemize}

We shall now present two criteria that allow to distinguish
Hamiltonian and symplectic structures between local solutions
of~\eqref{eq:bouss-new:51} and~\eqref{eq:bouss-new:52}.

\begin{theorem}
  \label{thm:gener-outl-comp-1}
  Let~$a=(a^1,\dots,a^m)$ be a solution
  of~\eqref{eq:bouss-new:51}\textup{,}
  where~$a^j=\sum_{l,i}a_i^{jl}\rho_i^l$. Consider the
  function~$A=\sum_ja^j\rho^j$. Then the corresponding
  $\mathcal{C}$-differential operator is a Hamiltonian structure on
  the equation~$\mathcal{E}$ if and only if
  \begin{equation}
    \label{eq:bouss-new:53}
    \sum_j\fd{A}{\rho^j}\rho^j=-2A
  \end{equation}
  and
  \begin{equation}
    \label{eq:bouss-new:54}
    \delta\sum_j\fd{A}{u^j}\cdot\fd{A}{\rho^j}=0.
  \end{equation}
  In addition\textup{,} two Hamiltonian structures with the
  functions~$A$ and~$A'$ are compatible if and only if
  \begin{equation}
    \label{eq:bouss-new:55}
    \delta\sum_j\Big(\fd{A}{u^j}\cdot\fd{A'}{\rho^j}+
    \fd{A'}{u^j}\cdot\fd{A}{\rho^j}\Big)=0.
  \end{equation}
\end{theorem}

\begin{theorem}
  \label{thm:gener-outl-comp-2}
  Let~$b=(b^1,\dots,b^m)$ be a solution
  of~\eqref{eq:bouss-new:52}\textup{,}
  where~$b^j=\sum_{l,i}b_i^{jl}\omega_i^l$. Consider the
  function~$B=\sum_jb^j\omega^j$. Then the corresponding
  $\mathcal{C}$-differential operator is a symplectic structure on the
  equation~$\mathcal{E}$ if and only if
  \begin{equation}
    \label{eq:bouss-new:56}
    \sum_j\fd{B}{\omega^j}\omega^j=-2B
  \end{equation}
  and
  \begin{equation}
    \label{eq:bouss-new:57}
    \delta\sum_j\fd{B}{u^j}\omega^j=0.
  \end{equation}
\end{theorem}

\begin{remark}
  To use these two results correctly, note that
  \begin{enumerate}
  \item The variables~$\rho_i^j$ and~$\omega_i^j$ should be considered
    as \emph{odd} ones.
  \item The operator~$\delta$ is the \emph{total} Euler operator in
    the corresponding covering equation, i.e., 
    \begin{equation*}
      \delta=
      (\fd{}{u^1},\dots,\fd{}{u^m},\fd{}{\rho^1},\dots,\fd{}{\rho^m})
    \end{equation*}
    for the $\ell_{\mathcal{E}}^*$-covering and
    \begin{equation*}
      \delta=
      (\fd{}{u^1},\dots,\fd{}{u^m},\fd{}{\omega^1},\dots,\fd{}{\omega^m})
    \end{equation*}
    for the $\ell_{\mathcal{E}}$-covering.
  \end{enumerate}
\end{remark}

\begin{remark}
  Conditions~\eqref{eq:bouss-new:53} and~\eqref{eq:bouss-new:56}
  guarantee that the corresponding operators are skew-adjoint.
  Conditions~\eqref{eq:bouss-new:54}, \eqref{eq:bouss-new:55},
  and~\eqref{eq:bouss-new:57} are equivalent
  to~\eqref{eq:bouss-new:26}, \eqref{eq:bouss-new:27},
  and~\eqref{eq:bouss-new:32}, respectively. An amazing property of
  evolution equations in ``general position'' was established
  in~\cite{Gessler} and~\cite{KKV-AAM} and is the following one.

\begin{theorem}
  \label{thm:gener-outl-comp-3}
  Let~$\mathcal{E}$ be an evolution equation of the
  form~\eqref{eq:bouss-new:2} and assume that it is not a first-order
  scalar equation. Then if the symbol of the right-hand side is
  nondegenerate, one has
  \begin{itemize}
  \item any skew-adjoint $\mathcal{C}$-differential operator
    satisfying~\eqref{eq:bouss-new:37}
    satisfies~\eqref{eq:bouss-new:54} as well \textup{(}any bivector
    is Poissonian\textup{)}\textup{,}
  \item any two skew-adjoint $\mathcal{C}$-differential operators
    satisfying~\eqref{eq:bouss-new:37}
    satisfy~\eqref{eq:bouss-new:55} as well \textup{(}any two
    Hamiltonian structures are compatible\textup{)}\textup{,}
  \item any skew-adjoint $\mathcal{C}$-differential operator
    satisfying~\eqref{eq:bouss-new:38}
    satisfies~\eqref{eq:bouss-new:57} as well \textup{(}any two-form
    is closed\textup{)}.
  \end{itemize}
\end{theorem}
\end{remark}

\section{Preparatory computations}
\label{sec:prep-comp}
We now come back to the dB-equation
\begin{equation*}
  w_t=u_x,\quad
  u_t=ww_x+v_x,\quad
  v_t=-uw_x-3wu_x
\end{equation*}
and first of all note that it is homogeneous with respect to the
following gradings (weights)
\begin{equation*}
  \abs{x}=-1,\ \abs{t}=-2,\ \abs{w}=2,\ \abs{u}=3,\ \abs{v}=4
\end{equation*}
and corresponding weights of all variables and their polynomial
functions. All constructions used in subsequent computations become
homogeneous and we can restrict ourselves to homogeneous components.

\subsection{Nonlocal variables}
\label{subsec:nonlocal-variables}

We looked for the coverings associated with conservation laws of the
dB-equation and found the variables~$p_1,p_2,p_3,p_5,\dots$, defined
by
\begin{align*}
  (p_1)_x & = w,                       &&(p_1)_t=u,\\
  (p_2)_x & = u,                       &&(p_2)_t=\frac{1}{2}(2v+w^2),\\
  (p_3)_x & = v + w^2,                 &&(p_3)_t=-uw,\\
  (p_5)_x & = (u^2 + 2 v w + 2 w^3)/2, &&(p_5)_t=uv.
\end{align*}
The subscript refers to the grading of the corresponding
variable. Thus,
\begin{equation*}
  \abs{p_1}=1,\ \abs{p_2}=2,\ \abs{p_3}=3,\ \abs{p_5}=5.
\end{equation*}
Further computations show that exactly one new conservation law arises
at each grading level except for gradings~$4,8,12,\dots$ We use the
corresponding nonlocal variables to construct nonlocal symmetries and
cosymmetries in Subsections~\ref{subsec:symmetries}
and~\ref{subsec:cosymmetries}.

\begin{remark}
  Void positions with gradings divisible by~$4$ are occupied by
  nonlocal variables of the next level of nonlocality.
  \begin{align*}
    (q_0)_x & = p_1,     &&(q_0)_t=p_2,\\
    (q_4)_x & = p_3w-3p_5,&&(q_4)_t=p_3u-2p_6,\\
  \end{align*}
  etc. Moreover, we found two independent nonlocal conservation laws
  for each of gradings~$8$ and~$12$.
\end{remark}

\subsection{Symmetries}
\label{subsec:symmetries}

The defining equations for symmetries (or their shadows in the
nonlocal case) of the dB-equation are
\begin{align}
  \label{eq:bouss-new:58}
  \tilde{D}_t(\phi^w)& =\tilde{D}_x (\phi^u), \nonumber \\
  \tilde{D}_t(\phi^u)& =w \tilde{D}_x(\phi^w) +
  w_1 \phi^w +\tilde{D}_x(\phi^v), \\
  \tilde{D}_t(\phi^v)& =-u \tilde{D}_x(\phi^w)-3 u_1 \phi^w
  -3 w \tilde{D}_x(\phi^u)- w_1  \phi^u. \nonumber
\end{align}
The grading of a symmetry is understood as the grading of the
corresponding vector field and hence
\begin{equation*}
  \abs{\phi}=\abs{\phi^w}-\abs{w}=\abs{\phi^u}
  -\abs{u}=\abs{\phi^v}-\abs{v}.\
\end{equation*}

\subsubsection{$(x,t)$-independent local symmetries}
\label{ssubsec:x-t-independent}

Solving equations~\eqref{eq:bouss-new:58} we found a number of local
symmetries independent of~$x$ and~$t$. Here are the results for
grading~$\le 7$:
\begin{align*}
  &\phi_{-4}^w=0, &&\phi_1^w = w_{1}, &&\phi_2^w = u_{1},\\
  &\phi_{-4}^u=0, &&\phi_1^u = u_{1}, &&\phi_2^u = w_{1} w + v_{1},\\
  &\phi_{-4}^v=1, &&\phi_1^v = v_{1}, &&\phi_2^v =  - w_{1} u - 3 u_{1} w,\\
\end{align*}
and
\begin{align*}
  \phi_3^w  & = 2 w_{1} w + v_{1},\\
  \phi_3^u  & =  - w_{1} u - u_{1} w,\\
  \phi_3^v  & =  - 3 w_{1} w^2 - u_{1} u - 2 v_{1} w,\\[1mm]
  \phi_5^w & = w_{1} (v + 3 w^2) + u_{1} u + v_{1} w, \\
  \phi_5^u & = u_{1} v + v_{1} u, \\
  \phi_5^v & = w_{1} ( - u^2 - 3 w^3) - 4 u_{1} u w + v_{1} (v - w^2),
  \\[1mm]
  \phi_6^w & = 2 w_{1} u w + u_{1} (2 v + w^2) + 2 v_{1} u, \\
  \phi_6^u & = w_{1} ( - 2 u^2 + 2 v w + w^3)
  - 4 u_{1} u w + v_{1} (2 v + w^2), \\
  \phi_6^v & = w_{1} u ( - 2 v - 7 w^2) + u_{1} ( - 2 u^2 - 6 v w - 3
  w^3)  - 6 v_{1} u w, \\[1mm]
  \phi_7^w & = w_{1} ( - 3 u^2 + 12 v w + 14 w^3)
  - 6 u_{1} u w + 6 v_{1} (v + w^2), \\
  \phi_7^u & = 6 w_{1} u ( - v - 2 w^2) + u_{1} ( - 3 u^2 - 6 v w - 4
  w^3)  - 6 v_{1} u w, \\
  \phi_7^v & = 6 w_{1} w (u^2 - 3 v w - 3 w^3) + 6 u_{1} u ( - v + 2
  w^2) + v_{1} ( - 3 u^2 - 12 v w - 10 w^3).
\end{align*}
The subscript in the notation above equals the grading of the
corresponding symmetry. We computed this kind of symmetries up to
grading~$15$ and found that they exist at all levels except for~$4$,
$8$, $12$.

\begin{remark}
  Note that only the first three symmetries are classical ones. All
  other symmetries are higher despite the fact that their jet order
  equals~$1$.
\end{remark}

\subsubsection{Nonlocal symmetries}
\label{ssubsec:nonlocal-symmetries}

Nonlocal symmetries (or, to be more precise, their shadows) arise at
gradings~$4$, $8$, $12$, etc. We computed these symmetries up to
grading~$16$ and found that at level~$4$ there exists one symmetry, at
level~$8$ two of them, at level~$12$ three symmetries arise, etc. For
example, at level~$4$ we have the symmetry~$\phi_{4,1}$ with the
components
\begin{align*}
  \phi_{4,1}^w & =  - 20p_3 w_{1} - 12p_2 u_{1}
  - 6p_1 (2 w_{1} w + v_{1}) + 9 u^2 + 16 v w + 16 w^3\\
  \phi_{4,1}^u & =  - 20p_3 u_{1} - 12p_2(w_{1} w +  v_{1})
  + 6p_1 (w_{1} u + u_{1} w) + 8 u (3 v + w^2),\\
  \phi_{4,1}^v & =  - 20 p_3 v_{1} + 12p_2 (w_{1} u
  + 3 u_{1} w) + 6 p_1 (3 w_{1} w^2 + u_{1} u + 2 v_{1} w)\\
  &- 39 u^2 w + 16 v^2 - 12 w^4,
\end{align*}
while at level~$8$ we have two symmetries~$\phi_{8,1}$
and~$\phi_{8,2}$ with the components
\begin{align*}
  \phi_{8,1}^w & =  - 1764 { p_{7}} w_{1} - 980 { p_{6}} u_{1}
  + 1176 { p_{5}} ( - 2 w_{1} w - v_{1}) \\
  &+ 280 { p_{3}} (w_{1} v + 3 w_{1} w^2 + u_{1} u + v_{1} w)
  + 42 {  p_{2}} (2 w_{1} u  + 2 u_{1} v + u_{1} w^2 + 2 v_{1} u) \\
  &+ 14 (54 u^2 v + 9 u^2 w^2 + 48 v^2 w + 96 v w^3 + 52 w^5),\\
  \phi_{8,1}^u & =  - 1764 { p_{7}} u_{1} - 980 { p_{6}} (w_{1} w
  + v_{1}) + 1176 { p_{5}} (w_{1} u + u_{1} w)
  + 280 { p_{3}} (u_{1} v + v_{1} u)\\
  & + 42 { p_{2}} ( - 2 w_{1} u^2 + 2 w_{1} v w + w_{1} w^3
  - 4 u_{1} u w + 2 v_{1} v + v_{1} w^2)\\
  & + 28 u ( - 27 u^2 w + 36 v^2 + 24 v w^2 - w^4),\\
  \phi_{8,1}^v & =  - 1764 { p_{7}} v_{1}
  + 980 { p_{6}} (w_{1} u + 3 u_{1} w)
  + 1176 { p_{5}} (3 w_{1} w^2 + u_{1} u + 2 v_{1} w)\\
  & + 280 { p_{3}} ( - w_{1} u^2 - 3 w_{1} w^3 - 4 u_{1} u w
  + v_{1} v - v_{1} w^2)\\
  & + 42 { p_{2}} ( - 2 w_{1} u v - 7 w_{1} u w^2
  - 2 u_{1} u^2 - 6 u_{1} v w - 3 u_{1} w^3 - 6 v_{1} u w)\\
  & + 7 ( - 27 u^4 - 468 u^2 v w - 190 u^2 w^3 + 64 v^3
  - 144 v w^4 - 96 w^6)
  \end{align*}
and
\begin{align*}
  \phi^w(8,2) & =  - 2184 { p_{7}} w_{1}
  - 1456 { p_{6}} u_{1} + 2184 { p_{5}} ( - 2 w_{1} w - v_{1}) \\
  &+ 1040 { p_{3}} (w_{1} v + 3 w_{1} w^2 + u_{1} u + v_{1} w)\\
  & + 312 { p_{2}} (2 w_{1} u w + 2 u_{1} v + u_{1} w^2 + 2 v_{1} u)\\
  & + 52 { p_{1}} ( - 3 w_{1} u^2 + 12 w_{1} v w + 14 w_{1} w^3
  - 6 u_{1} u w + 6 v_{1} v + 6 v_{1} w^2),\\
  \phi^u(8,2) & =  - 2184 { p_{7}} u_{1} - 1456 { p_{6}} (w_{1} w
  + v_{1}) + 2184 { p_{5}} (w_{1} u + u_{1} w) \\
  &+ 1040 { p_{3}} (u_{1} v + v_{1} u)\\
  & + 312 { p_{2}} ( - 2 w_{1} u^2 + 2 w_{1} v w + w_{1} w^3
  - 4 u_{1} u w + 2 v_{1} v + v_{1} w^2)\\
  & + 52 { p_{1}} ( - 6 w_{1} u v - 12 w_{1} u w^2
  - 3 u_{1} u^2 - 6 u_{1} v w - 4 u_{1} w^3 - 6 v_{1} u w),\\
  \phi^v(8,2) & =  - 2184 { p_{7}} v_{1}
  + 1456 { p_{6}} (w_{1} u + 3 u_{1} w)
  + 2184 { p_{5}} (3 w_{1} w^2 + u_{1} u + 2 v_{1} w)\\
  & + 1040 { p_{3}} ( - w_{1} u^2 - 3 w_{1} w^3 - 4 u_{1} u w
  + v_{1} v - v_{1} w^2)\\
  & + 312 { p_{2}} ( - 2 w_{1} u v - 7 w_{1} u w^2
  - 2 u_{1} u^2 - 6 u_{1} v w - 3 u_{1} w^3 - 6 v_{1} u w)\\
  & + 52 { p_{1}} (6 w_{1} u^2 w - 18 w_{1} v w^2 - 18 w_{1} w^4 - 6
  u_{1} u v \\
  &+ 12 u_{1} u w^2 - 3 v_{1} u^2 - 12 v_{1} v w - 10 v_{1} w^3).
\end{align*}

\subsubsection{$(x,t)$-dependent symmetries}
\label{ssubsec:x-t-dependent}

These symmetries arise at levels~$0$, $4$, $8$, etc., and at each
level we have two new symmetries~$\bar{\phi}_{0,1}$, $\bar{\phi}_{0,2}$,
$\bar{\phi}_{4,1}$, $\bar{\phi}_{4,2},\dots$ For example, we have
\begin{align*}
  &\bar{\phi}_{0,1}^w=xw_1-2w, &&\bar{\phi}_{0,2}^w=tu_1+2w,\\
  &\bar{\phi}_{0,1}^u=xu_1-3u, &&\bar{\phi}_{0,2}^u=t(w_1w+v_1)+3u,\\
  &\bar{\phi}_{0,1}^v=xv_1-4v, &&\bar{\phi}_{0,2}^v=-t(w_1u+3u_1w)+4v
\end{align*}
and
\begin{align*}
  \bar{\phi}_{4,1}^w&=x(w_1v+3w_1w^2+u_1u+v_1w)-3p_1(2w_1w+v_1)
  -4p_2u_1-5p_3w_1,\\
  \bar{\phi}_{4,1}^u&=x(u_1v+v_1u)+3p_1(w_1u+u_1w)-4p_2(w_1w+v_1)
  -5p_3u_1,\\
  \bar{\phi}_{4,1}^v&=-x(w_1u^2+3w_1w^3+4u_1uw-v_1v+v_1w^2)
  +3p_1(3w_1w^2+u_1u+2v_1w)\\
  &+4p_2(w_1u+3u_1w)-5p_3v_1,\\[1mm]
  \bar{\phi}_{4,2}^w&=t(2w_1uw+2u_1v+u_1w^2+2v_1u)
  +4p_1(2w_1w+v_1)+6p_2u_1+8p_3w_1,\\
  \bar{\phi}_{4,2}^u&=-t(2w_1u^2-2w_1vw-w_1w^3+4u_1uw-2v_1v-v_1w^2)\\
  &-4p_1(w_1u+u_1w)+6p_2(w_1w+v_1)+8p_3u_1,\\
  \bar{\phi}_{4,2}^v&=-t(2w_1uv+7w_1uw^2+2u_1u^2+6u_1vw+3u_1w^3+6v_1uw)\\
  &-4p_1(3w_1w^2+u_1u+2v_1w)-6p_2(w_1u+3u_1w)+8p_3v_1.
\end{align*}
Note that only the two symmetries are local.

\subsubsection{Distribution of symmetries}
\label{ssubsec:distr-symm}

We present the distribution of symmetries described above in
Table~\ref{tab:sym-distr}.

\begin{table}[h]
  \centering\tiny
  \begin{tabular}{ccccccccccccccccc}
    \hline
    -4&-3&-2&-1&0&1&2&3&4&5&6&7&8&9&10&11&12\\
    \hline
    $\phi_{-4}$&&&&&$\phi_1$&$\phi_2$&$\phi_3$&&$\phi_5$&$\phi_6$&$\phi_7$&&$\phi_9$&$\phi_{10}$&$\phi_{11}$&\\
    &&&&&&&&$\phi_{4,1}$&&&&$\phi_{8,1}$&&&&$\phi_{12,1}$\\
    &&&&&&&&&&&&$\phi_{8,2}$&&&&$\phi_{12,2}$\\
    &&&&&&&&&&&&&&&&$\phi_{12,3}$\\
    &&&&$\bar{\phi}_{0,1}$&&&&$\bar{\phi}_{4,1}$&&&&$\bar{\phi}_{8,1}$&&&&$\bar{\phi}_{12,1}$\\
    &&&&$\bar{\phi}_{0,2}$&&&&$\bar{\phi}_{4,2}$&&&&$\bar{\phi}_{8,2}$&&&&$\bar{\phi}_{12,2}$\\
    \hline
  \end{tabular}\normalsize
  \caption{Distribution of symmetries along gradings}
  \label{tab:sym-distr}
\end{table}

\subsection{Cosymmetries}
\label{subsec:cosymmetries}

The defining equations for cosymmetries (or their shadows) of the
dB-equation are

\begin{align}
  \label{eq:bouss-new:59}
  \tilde{D_t}(\psi^w) & = w \tilde{D_x}(\psi^u)
  -u \tilde{D_x}(\psi^v)+2 u_{1} \psi^v,\nonumber\\
  \tilde{D_t}(\psi^u) & =   \tilde{D_x}(\psi^w)
  -3 w \tilde{D_x}(\psi^v)-2 w_{1} \psi^v,\\
  \tilde{D_t}(\psi^v) & = \tilde{D_x}(\psi^u). \nonumber
\end{align}

The grading of a cosymmetry is understood as the grading of the
corresponding differential form and thus
\begin{equation*}
  \abs{\psi}=\abs{\psi^w}+\abs{w}=\abs{\psi^u}+\abs{u}=
  \abs{\psi^v}+\abs{v}.
\end{equation*}

\subsubsection{$(x,t)$-independent local cosymmetries}
\label{ssubsec:x-t-independent-co}
We computed these cosymmetries up to grading~$16$ and found that they
exist at all levels except for~$1$, $5$, $9$, etc. Here we present the
results for gradings~$\le 8$:
\begin{align*}
  &\psi_2^w = 1,&&\psi_3^w = 0,&&\psi_4^w = w, \\
  &\psi_2^u = 0,&&\psi_3^u = 1,&&\psi_4^u = 0, \\
  &\psi_2^v = 0,&&\psi_3^v = 0,&&\psi_4^v = \frac{1}{2}, \\[1mm]
  &\psi_6^w = v + 3 w^2,&&\psi_7^w = u w,
  &&\psi_8^w = \frac{1}{14}( - 3 u^2 + 12 v w + 14 w^3)\\
  &\psi_6^u = u,&&\psi_7^u = \frac{1}{2}(2 v + w^2),
  &&\psi_8^u = -\frac{3}{7} u w,\\
  &\psi_6^v = w,&&\psi_7^v = u,&&\psi_8^v = \frac{3}{7}(v + w^2).
\end{align*}

\subsubsection{Nonlocal cosymmetries}
\label{ssubsec:nonl-cosymm}
Nonlocal cosymmetries independent of~$x$ and~$t$ were found at
levels~$9$ (one cosymmetry), $13$ (two cosymmetries), and~$17$ (three
cosymmetries). For example, we have
\begin{align*}
  \psi_{9,1}^w&=\frac{1}{14}(-42p_7-84p_5w+20p_3(v+3w^2)+12p_2uw\\
  &+p_1(-3u^2+12vw+14w^3)),\\
  \psi_{9,1}^u&=\frac{1}{7}(-14p_6+10p_3u+3p_2(2v+w^2)-3p_1uw),\\
  \psi_{9,1}^v&=\frac{1}{7}(-21p_5+10p_3w+6p_2u+3p_1(v+w^2)),\\[1mm]
  \psi_{13,1}^w&=\frac{1}{3}(195p_{11}+484p_9w-126p_7(v-3w^2)
  -70p_6uw\\
  &+14p_5(3u^2-12vw-14w^3)+5p_3(2v^2+12vw^2+11w^4)\\
  &+p_2u(-2u^2+6vw+3w^3)),\\
  \psi_{13,1}^u&=\frac{1}{12}(297p_{10}-504p_7u-140p_6(2v+w^2)
  +336p_5uw+80p_3uv\\
  &+3p_2(-8u^2w+4v^2+4vw^2+w^4)),\\
  \psi_{13,1}^v&=\frac{1}{3}(242p_9-126p_7w-70p_6u-84p_5(v+w^2)
  +10p_3(u^2+2vw+2w^3)\\
  &+3p_2u(2v+w^2)),\\
  \intertext{and}
  \psi_{13,2}^w&=\frac{1}{9}(-351p_{11}-847p_9w+210p_7(v+3w^2)
  +112p_6uw\\
  &+21p_5(-3u^2+12vw+14w^3)-5p_3(2v^2+12vw^2+11w^4)\\
  &+p_1(-3u^2v-6u^2w^2+6v^2w+14vw^3+9w^5)),\\
  \psi_{13,2}^u&=\frac{1}{9}(-132p_{10}+210p_7u+56p_6(2v+w^2)
  -126p_5uw-20p_3uv\\
  &-p_1u(u^2+6vw+4w^3)),\\
  \psi_{13,2}^v&=\frac{1}{18}(-847p_9+420p_7w+224p_6u+252p_5(v+w^2)\\
  &-20p_3(u^2+2vw+2w^3)+p_1(-6u^2w+6v^2+12vw^2+7w^4)).
\end{align*}

\subsubsection{$(x,t)$-dependent cosymmetries}
\label{ssubsec:x-t-dependentco}

These cosymmetries exist at levels~$1$, $5$, $9$, etc. The first one,
\begin{equation*}
  \bar{\psi}_{1,1}^w=x,\qquad\bar{\psi}_{1,1}^u=t,\qquad
  \bar{\psi}_{1,1}^v=0,
\end{equation*}
is local, all others are nonlocal: two at level~$5$,
\begin{align*}
  &\bar{\psi}_{5,1}^w=x(v+3w^2)-6p_1w-5p_3,
  &&\bar{\psi}_{5,2}^w=tuw+4p_1w+4p_3,\\
  &\bar{\psi}_{5,1}^u=xu-4p_2,
  &&\bar{\psi}_{5,2}^u=\frac{1}{2}(t(2v+w^2)+6p_2),\\
  &\bar{\psi}_{5,1}^v=xw-3p_1,
  &&\bar{\psi}_{5,2}^v=tu+2p_1,
\end{align*}
two at level~$9$,
\begin{align*}
  \bar{\psi}_{9,1}^w&=\frac{1}{11}(x(2v^2+12vw^2+11w^4)+8p_2uw
  +20p_3(v+3w^2)\\
  &-112p_5w-63p_7),\\
  \bar{\psi}_{9,1}^u&=\frac{4}{11}(xuv+p_2(2v+w^2)+5p_3-10p_6),\\
  \bar{\psi}_{9,1}^v&=\frac{2}{11}(x(u^2+2vw+2w^3)
  +4p_2u+10p_3w-28p_5),\\[1mm]
  \bar{\psi}_{9,2}^w&=\frac{1}{3}(tu(-2u^2+6vw+3w^3)-6p_2uw
  -16p_3(v+3w^2)+96p_5w+56p_7),\\
  \bar{\psi}_{9,2}^u&=\frac{1}{12}(3t(-8u^2w+4v^2+4vw^2+w^4)
  -12p_2(2v+w^2)-64p_3u+140p_6),\\
  \bar{\psi}_{9,2}^w&=\frac{1}{3}(3tu(2v+w^2)-6p_2u-16p_3w+48p_5),
\end{align*}
etc.

\subsubsection{Distribution of cosymmetries}
\label{ssubsec:distr-cosymm}

We present the distribution of cosymmetries described above in
Table~\ref{tab:cosym-distr}.

\begin{table}[h]
  \centering\tiny
  \begin{tabular}{ccccccccccccccccc}
    \hline
    1&2&3&4&5&6&7&8&9&10&11&12&13&14&15&16&17\\
    \hline
    &$\psi_2$&$\psi_3$&$\psi_4$&&$\psi_6$&$\psi_7$&$\psi_8$&&$\psi_{10}$&$\psi_{11}$&$\psi_{12}$&&$\psi_{14}$&$\psi_{15}$&$\psi_{16}$&\\
    &&&&&&&&$\psi_{9,1}$&&&&$\psi_{13,1}$&&&&$\psi_{17,1}$\\
    &&&&&&&&&&&&$\psi_{13,2}$&&&&$\psi_{17,2}$\\
    &&&&&&&&&&&&&&&&$\psi_{17,3}$\\
    $\bar{\psi}_{1,1}$&&&&$\bar{\psi}_{5,1}$&&&&$\bar{\psi}_{9,1}$&&&&$\bar{\psi}_{13,1}$&&&&$\bar{\psi}_{17,1}$\\
    &&&&$\bar{\psi}_{5,2}$&&&&$\bar{\psi}_{9,2}$&&&&$\bar{\psi}_{13,2}$&&&&$\bar{\psi}_{17,2}$\\
    \hline
  \end{tabular}\normalsize
  \caption{Distribution of cosymmetries along gradings}
  \label{tab:cosym-distr}
\end{table}

\subsection{Nonlocal vectors}
\label{subsec:nonlocal-vectors}

We construct nonlocal vectors on the $\ell_{\mathcal{E}}^*$-covering
(see Remark~\ref{rem:nonlocal-vect-covect}) associated with
symmetries~$\phi_{-4},\dots,\phi_7$ and denote them
by~$\rho^{-4},\dots,\rho^7$, respectively. Thus, by definition, we
have
\begin{equation*}
  \tilde{D}_x(\rho^i)=\rho^w\phi_i^w+\rho^u\phi_i^u+\rho^v\phi_i^v,\qquad
  i=-4,1,2,3,5,6,7.
\end{equation*}
The corresponding $t$-components are given by the following relations:
\begin{equation*}
  \tilde{D}_t(\rho^i)=\rho^w\phi_i^u+\rho^u(\phi_i^v+\phi_i^ww)
  -\rho^v(3\phi_i^uw+\phi_i^wu).
\end{equation*}
Here~$\rho^w$, $\rho^u$, $\rho^v$ are the coordinates in the
$\ell_{\mathcal{E}}^*$-covering (see equations~\eqref{eq:bouss-new:61}
below).

\subsection{Nonlocal forms}
\label{subsec:nonlocal-forms}

We also constructed nonlocal forms associated to the
cosymmetries~$\psi_2$, $\psi_3$, $\psi_4$, and~$\psi_6$, $\psi_7$,
$\psi_8$. We denote these forms by~$\omega^2,\dots,\omega^8$ and have
by definition
\begin{equation*}
  \tilde{D}_x(\omega^i)=\psi_i^w\omega^w+\psi_i^u\omega^u
  +\psi_i^v\omega^v,\qquad i=2,3,4,6,7,8,
\end{equation*}
where~$\omega^w$, $\omega^u$, $\omega^v$ are the coordinates in the
$\ell_{\mathcal{E}}$-covering (see
equations~\eqref{eq:bouss-new:60}).

The $t$-components of the nonlocal forms under consideration are given
by the equations
\begin{equation*}
  \tilde{D}_t(\omega^i)=(\psi_i^uw-\psi_i^vu)\omega^w
  +(\psi_i^w-3\psi_i^vw)\omega^u+\psi_i^u\omega^v.
\end{equation*}

\section{The main results}
\label{sec:main-results}
We begin with the defining equations for the $\ell_{\mathcal{E}}$- and
$\ell_{\mathcal{E}}^*$-coverings of the dB-equation. In the first case
we have
\begin{align}\label{eq:bouss-new:60}\nonumber
  \omega_t^w&=\omega_x^u,\\
  \omega_t^u&=w\omega_x^w+w_x\omega^w+\omega_x^v,\\\nonumber
  \omega_t^v&=-u\omega_x^w-w_x\omega^u-3w\omega_x^u-3u_x\omega^w,
\end{align}
while the equations of the $\ell_{\mathcal{E}}^*$-covering are

\begin{align}\label{eq:bouss-new:61}\nonumber
  \rho_t^w&=w\rho_x^u-u\rho_x^v+2u_x\rho^v,\\
  \rho_t^u&=\rho_x^w-3w\rho_x^v-2w_x\rho^v,\\\nonumber
  \rho_t^v&=\rho_x^u.
\end{align}

The total derivatives on the $\ell_{\mathcal{E}}$-covering are of the
form
\begin{align}\label{eq:bouss-new:62}\nonumber
  \tilde{D}_x&=D_x+\sum_{i\ge0}\Big(\omega_{i+1}^w\pd{}{\omega_i^w}+
  \omega_{i+1}^u\pd{}{\omega_i^u}+\omega_{i+1}^v\pd{}{\omega_i^v}\Big),\\
  \tilde{D}_t&=D_t+\sum_{i\ge0}\Big(\omega_{i+1}^u\pd{}{\omega_i^w}+
  \tilde{D}_x^i(w\omega_1^w+w_1\omega^w+\omega_1^v)\pd{}{\omega_i^u}\\
  \nonumber
  &-\tilde{D}_x^i(u\omega_1^w+w_1\omega^u+3w\omega_1^u
  +3u_1\omega^w)\pd{}{\omega_i^v}\Big),
\end{align}
where~$D_x$ and~$D_t$ are total derivatives on the dB-equation, while
on the $\ell_{\mathcal{E}}^*$-covering they are presented by
\begin{align}\label{eq:bouss-new:63}\nonumber
  \tilde{D}_x&=D_x+\sum_{i\ge0}\Big(\rho_{i+1}^w\pd{}{\rho_i^w}+
  \rho_{i+1}^u\pd{}{\rho_i^u}+\rho_{i+1}^v\pd{}{\rho_i^v}\Big),\\
  \tilde{D}_t&=D_t+\sum_{\ge0}\Big(\tilde{D}_x(w\rho_1^u-u\rho_1^v
  +2u_1\rho^v)\pd{}{\rho_i^w}\\
  &+\tilde{D}_x(\rho_1^w-3w\rho_1^v-2w_1\rho^v)\pd{}{\rho_i^u}
  +\rho_{i+1}^u\pd{}{\rho_i^v}\Big).\nonumber
\end{align}

\subsection{Recursion operators for symmetries}
\label{subsec:recurs-oper-symm}

We construct recursion operators
$\mathcal{R}=(\mathcal{R}^w,\mathcal{R}^u,\mathcal{R}^v)$ for
symmetries of the dB-equation solving the system
\begin{align*}
  \tilde{D}_t(\mathcal{R}^w)& = \tilde{D}_x (\mathcal{R}^u), \\
  \tilde{D}_t(\mathcal{R}^u)& = w \tilde{D}_x(\mathcal{R}^w)
  + w_{1} \mathcal{R}^w + \tilde{D}_x(\mathcal{R}^v), \\
  \tilde{D}_t(\mathcal{R}^v) & =  - u \tilde{D}_x(\mathcal{R}^w)
  - 3 u_{1} \mathcal{R}^w -3 w \tilde{D}_x(\mathcal{R}^u)
  - w_{1}  \mathcal{R}^u,
\end{align*}
where the total derivatives are defined by
formulas~\eqref{eq:bouss-new:62}. Recall that we look for solutions
linear in~$\omega_i$'s and admit their dependency on nonlocal
forms. Besides the trivial solution
\begin{equation*}
  \mathcal{R}_0^w  = \omega^w, \quad
  \mathcal{R}_0^u  = \omega^u, \quad
  \mathcal{R}_0^v  = \omega^v
\end{equation*}
that corresponds to the identity operator, we found the following two
nontrivial solutions of grading~$4$:
\begin{align*}
  \mathcal{R}_{4,1}^w & = (2 \omega_4 w_1 + \omega^3 u_1
  + \omega^{2} (2 w_1 w + v_1))/2, \\
  \mathcal{R}_{4,1}^u & = (2 \omega^4 u_1 + \omega^3 (w_1 w + v_1)
  - \omega^{2} (w_1 u + u_1 w))/2, \\
  \mathcal{R}_{4,1}^v & = (2 \omega^4 v_1
  + \omega^3 ( - w_1 u - 3 u_1 w)
  + \omega^{2} ( - 3 w_1 w^2 - u_1 u - 2 v_1 w))/2, \\[1mm]
  \mathcal{R}_{4,2}^w & = ( - 4 \omega^4 w_1 - \omega^3 u_1
  + 2 \omega^v w + 3 \omega^u u + 4 \omega^w (v + 2 w^2))/8, \\
  \mathcal{R}_{4, 2}^u & = ( - 4 \omega^4 u_1
  - \omega^3 (w_1 w + v_1) + 3 \omega^v u + 2 \omega^u (2 v + w^2)
  + \omega^w u w)/8, \\
  \mathcal{R}_{4,2}^v & = ( - 4 \omega^4 v_1 + \omega^3 (w_1 u
  + 3 u_1 w) + 4 \omega^v v - 11 \omega^u u w
  + 3 \omega^w ( - u^2 - 2 w^3))/8.
\end{align*}

\begin{remark}
  We also found three solutions at level~$8$, but they are too
  cumbersome to present them here.
\end{remark}

The operators corresponding to the above solutions are
\begin{equation}
  \label{eq:bouss-new:64}
  \mathcal{R}_{4,1}=\frac{1}{2}
  \langle2\phi_1,\phi_2,\phi_3\mid D_x^{-1}\mid\psi_4,\psi_3,\psi_2\rangle
\end{equation}
and
\begin{multline}
  \label{eq:bouss-new:65}
  \mathcal{R}_{4,2}=-\frac{1}{8}\langle4\phi_1,\phi_2\mid
  D_x^{-1}\mid\psi_4,\psi_3\rangle
  \\
  +\frac{1}{8}
  \begin{pmatrix}
    4(v + 2 w^2)&3u&2w\\
    u w&2(2 v + w^2)&3u \\
    -3(u^2+2 w^3)&- 11u w&4v
  \end{pmatrix}
\end{multline}
(here and below we use the notation introduced in
Remark~\ref{rem:nonlocal-theory-notation}).  The explicit
expressions for symmetries~$\phi_i$ and cosymmetries~$\psi_j$ are
given in Subsections~\ref{ssubsec:x-t-independent}
and~\ref{ssubsec:x-t-independent-co}.

\subsection{Recursion operators for cosymmetries}
\label{subsec:recurs-oper-cosymm}

Recursion operators for cosymmetries are of the
form~$\mathcal{R}^*=(\mathcal{R}^{*,u},\mathcal{R}^{*,u},\mathcal{R}^{*,v})$
and can be constructed from the equation
\begin{align*}
  \tilde{D}_t(\mathcal{R}^{*,w}) & = w \tilde{D}_x(\mathcal{R}^{*,u})
  -u \tilde{D}_x(\mathcal{R}^{*,v})+2 u_{1} \mathcal{R}^{*,v},\\
  \tilde{D}_t(\mathcal{R}^{*,u}) & = \tilde{D}_x(\mathcal{R}^{*,w})
  -3 w \tilde{D}_x(\mathcal{R}^{*,v})-2 w_{1} \mathcal{R}^{*,v},\\
  \tilde{D}_t(\mathcal{R}^{*,v}) & = \tilde{D}_x(\mathcal{R}^{*,u}),
\end{align*}
where the total derivatives are given by
formulas~\eqref{eq:bouss-new:63}. The first two nontrivial solutions
(both of grading~$4$) are
\begin{align*}
  R_{4,1}^{*,w} & = (\rho^{3} + 2 \rho^{1} w)/2, \\
  R_{4,1}^{*,u} & = \rho^{2}/2, \\
  R_{4,1}^{*,v} & = \rho^{1}/2, \\[1mm]
  R_{4,2}^{*,w} & = ( - 2 \rho^{3}
  + 3 \rho^v ( - 2 w^3 - u^2) + \rho^u w u + 4 \rho^w (2 w^2 + v))/8, \\
  R_{4,2}^{*,u} & = ( - \rho^{2} - 11 \rho^v w u
  + 2 \rho^u (w^2 + 2 v) + 3 \rho^w u)/8, \\
  R_{4,2}^{*,v} & = (4 \rho^v v + 3 \rho^u u + 2 \rho^w w)/8.
\end{align*}
The operators corresponding to these solutions are
\begin{equation}
  \label{eq:bouss-new:66}
  \mathcal{R}_{4,1}^*=\frac{1}{2}
  \langle2\psi_4,\psi_3,\psi_2\mid D_x^{-1}\mid\phi_1,\phi_2,\phi_3\rangle
\end{equation}
and
\begin{multline}
  \label{eq:bouss-new:67}
  \mathcal{R}_{4,2}^*=-\frac{1}{8}
  \langle\psi_3,2\psi_2\mid D_x^{-1}\mid\phi_2,\phi_3\rangle
  \\
  +\frac{1}{8}
  \begin{pmatrix}
    4(2 w^2 + v)&w u&-3(2 w^3+u^2)\\
    3u&2(w^2 + 2 v)&- 11w u\\
    2 w&3 u &4v
  \end{pmatrix}.
\end{multline}

\subsection{Hamiltonian structures}
\label{subsec:hamilt-struct}

The first step to construct Hamiltonian structures
$\mathcal{H}=(\mathcal{H}^w,\mathcal{H}^u,\mathcal{H}^v)$ is to solve
the equation
\begin{align*}
  \tilde{D}_t(\mathcal{H}^w)& = \tilde{D}_x (\mathcal{H}^u), \\
  \tilde{D}_t(\mathcal{H}^u)& = w \tilde{D}_x(\mathcal{H}^w)
  + w_{1} \mathcal{H}^w + \tilde{D}_x(\mathcal{H}^v), \\
  \tilde{D}_t(\mathcal{H}^v) & =  - u \tilde{D}_x(\mathcal{H}^w)
  - 3 u_{1} \mathcal{H}^w -3 w \tilde{D}_x(\mathcal{H}^u)
  - w_{1}  \mathcal{H}^u
\end{align*}
with total derivatives defined by
formulas~\eqref{eq:bouss-new:63}. The first three solutions are
\begin{align*}
  \mathcal{H}_{0,1}^w & = \rho_1^v, \\
  \mathcal{H}_{0,1}^u & = \rho_1^u, \\
  \mathcal{H}_{0,1}^v & = -4\rho_1^v w + \rho_1^w - 2 \rho^v w_{1}, \\[1mm]
  \mathcal{H}_{4,1}^w & = (4 \rho_1^v v + 3 \rho_1^u u + 2 \rho_1^w w
  + 2 \rho^v v_{1} + \rho^u u_{1})/2, \\
  \mathcal{H}_{4,1}^u & = (- 11\rho_1^v w u + 2\rho_1^u(w^2 + 2 v)
  + 3 \rho_1^w u -2 \rho^v (w u_{1} + 4 u w_{1})\\
  & + \rho^u (w w_{1} + v_{1}))/2, \\
  \mathcal{H}_{4,1}^v & = (-\rho_1^v(6 w^3 + 16 w v + 3 u^2)
  - 11 \rho_1^u w u + 4 \rho_1^w v\\
  & - 2 \rho^v (3 w^2 w_{1} + 2 w v_{1} + u u_{1} + 4 v w_{1})\\
  & + \rho^u ( - 3 w u_{1} - u w_{1}))/2, \\[1mm]
  \mathcal{H}_{4,2}^w & = \rho^v v_{1} + \rho^u u_{1} + \rho^w w_{1}, \\
  \mathcal{H}_{4,2}^u & = \rho^v (3 w u_{1} + u w_{1})
  + \rho^u (w w_{1} + v_{1}) + \rho^w u_{1}, \\
  \mathcal{H}_{4, 2}^v & = \rho^v ( - 3 w^2 w_{1} - 4 w v_{1}
  - u u_{1}) + \rho^u ( - 3 w u_{1} - u w_{1}) + \rho^w v_{1}.
\end{align*}
The operators corresponding to these solutions are
\begin{equation*}
  \mathcal{H}_{0,1}=\mathcal{H}_1=
  \begin{pmatrix}
    0&0&D_x\\
    0&D_x&0\\
    D_x &0&-4D_x w - 2  w_1
  \end{pmatrix}
\end{equation*}
and
\begin{align*}
  \mathcal{H}_{4,1}&=\frac{1}{2}
  \begin{pmatrix}
    2wD_x&3uD_x+u_1&2(2vD_x+v_1)\\
    3uD_x&2(w^2+2v)D_x+ww_1+v_1&-11wuD_x-2(wu_1+4uw_1)\\
    4vD_x&-11wuD_x-3wu_1-uw_1&h_{2,2}^1D_x+h_{2,2}^0
  \end{pmatrix},\\
  \mathcal{H}_{4,2}&=
  \begin{pmatrix}
    w_1&u_1&v_1\\
    u_1&ww_1+v1&-3wu_1-uw_1\\
    v_1&-3wu_1-uw_1&-3w^2w_1-4wv_1-uu_1
  \end{pmatrix},
\end{align*}
where
\begin{equation*}
  h_{2,2}^1=-(6w^3+16wv+3u^2),\qquad
  h_{2,2}^0=-2(3w^2w_1+2wv_1+uu_1+4vw_1).
\end{equation*}

The operator~$\mathcal{H}_1$ is skew-adjoint and by
Theorem~\ref{thm:gener-outl-comp-3} is a Hamiltonian structure for
the dB-equation, but neither of the last two operators is
Hamiltonian. Nevertheless, their linear combination
\begin{equation*}
  \mathcal{H}_2=\mathcal{H}_{4,1}+\frac{1}{2}\mathcal{H}_{4,2}
\end{equation*}
is skew-adjoint and consequently Hamiltonian. Again, by
Theorem~\ref{thm:gener-outl-comp-3} the structures~$\mathcal{H}_1$
and~$\mathcal{H}_2$ are compatible.

\subsection{Symplectic structures}
\label{subsec:sympl-struct}

To find symplectic structures
$\mathcal{S}=(\mathcal{S}^w,\mathcal{S}^u,\mathcal{S})^v$, we solve
the equation
\begin{align*}
  \tilde{D}_t(\mathcal{S}^{w}) & = w \tilde{D}_x(\mathcal{S}^{u})
  -u \tilde{D}_x(\mathcal{S}^{v})+2 u_{1} \mathcal{S}^{v},\\
  \tilde{D}_t(\mathcal{S}^{u}) & = \tilde{D}_x(\mathcal{S}^{w})
  -3 w \tilde{D}_x(\mathcal{S}^{v})-2 w_{1} \mathcal{S}^{v},\\
  \tilde{D}_t(\mathcal{S}^{v}) & = \tilde{D}_x(\mathcal{S}^{u}),
\end{align*}
where the total derivatives are given by
formulas~\eqref{eq:bouss-new:62}. We obtained one solution
\begin{align*}
  \mathcal{S}_{0,1}^w & = \omega^{4} + \omega^{2} w, \\
  \mathcal{S}_{0,1}^u & = \omega^{3}/2, \\
  \mathcal{S}_{0,1}^v & = \omega^{2}/2
\end{align*}
of grading~$0$ and to solutions of grading~$4$:
\begin{align*}
  \mathcal{S}_{4,1}^w &=-14\omega^8-8\omega^6 w+4\omega^4(3w^2+v)+\omega^3wu,\\
  \mathcal{S}_{4,1}^u &=(-10\omega^7+8\omega^4 u+\omega^3(w^2 + 2 v))/2, \\
  \mathcal{S}_{4,1}^v &= -4\omega^6+4\omega^4 w+\omega^3 u, \\[1mm]
  \mathcal{S}_{4,2}^w &=(70\omega^8+36\omega^6 w-12\omega^4(3 w^2 + v)
  + \omega^{2} (14 w^3 + 12 w v - 3 u^2))/14, \\
  \mathcal{S}_{4,2}^u & =(12\omega^7-6\omega^4 u - 3 \omega^2 w u)/7, \\
  \mathcal{S}_{4,2}^v & =(9 \omega^6-6 \omega^4 w + 3 \omega^2 (w^2 + v))/7.
\end{align*}
The operator
\begin{equation*}
  \mathcal{S}_1=\mathcal{S}_{0,1}=
  \langle\psi_4,\frac{1}{2}\psi_3,\psi_2\mid D_x^{-1}\mid
  \psi_2,\psi_3,\psi_4\rangle
\end{equation*}
corresponds to the first solution, while to the second and the third
ones the operators
\begin{equation*}
  \mathcal{S}_{4,1}=
  \langle\psi_7,4\psi_6,-8\psi_4,-5\psi_3,-14\psi_2\mid D_x^{-1}\mid
  \psi_3,\psi_4,\psi_6,\psi_7,\psi_8\rangle
\end{equation*}
and
\begin{equation*}
  \mathcal{S}_{4,2}=
  \langle\psi_8,-\frac{6}{7}\psi_6,\frac{18}{7}\psi_4,
  \frac{12}{7}\psi_3,5\psi_2\mid D_x^{-1}\mid\psi_2,
  \psi_4,\psi_6,\psi_7,\psi_8\rangle
\end{equation*}
correspond.

The operator~$\mathcal{S}_1$ is skew-adjoint and hence is a symplectic
structure for the dB-equation. In addition, the operator
\begin{equation*}
  \mathcal{S}_2=2\mathcal{S}_{4,1}+7\mathcal{S}_{4,2}
\end{equation*}
is a symplectic structure as well.

\section{Interrelations}
\label{sec:interrelations}

We shall now present the basic algebraic relations between the above
described structures. They include description of the following
compositions:
\begin{equation*}
  \mathcal{R}\circ\mathcal{R},\ \mathcal{R}\circ\mathcal{H},\
  \mathcal{S}\circ\mathcal{R},\ \mathcal{H}\circ\mathcal{S},\
  \mathcal{S}\circ\mathcal{H},\ \mathcal{H}\circ\mathcal{R}^*,\
  \mathcal{R}^*\circ\mathcal{S},\ \mathcal{R}^*\circ\mathcal{R}^*,
\end{equation*}
together with the action of the operators~$\mathcal{R}$ on symmetries
and of~$\mathcal{R}^*$ on cosymmetries and as well as commutator
relations between symmetries.

Of course, some of these relations are deducible from the other ones,
but we prefer explicit presentation.

\subsection{$\mathcal{R}\circ\mathcal{R}$}
\label{subsec:rr}
Compositions of the first two recursion operators look as follows
\begin {align*}
R_{4,1}\circ R_{4,1} &=\frac{1}{4}R_{8,1}+\frac{7}{12}R_{8,2},\\
R_{4,1}\circ R_{4,2} &=-\frac{1}{16}R_{8,1},\\
R_{4,2}\circ R_{4,1} &=\frac{3}{16}R_{8,1}+\frac{7}{12}R_{8,2},\\
R_{4,2}\circ R_{4,2} &=-\frac{3}{64}R_{8,1}+\frac{13}{16}R_{8,3}.
\end{align*}
So we have
\begin{equation*}
  R_{4,2}\circ R_{4,1}=R_{4,1}\circ R_{4,2}+R_{4,1}\circ R_{4,1},
\end{equation*}
or
\begin{equation}
  \label{eq:bouss-new:68}
  [R_{4,2},R_{4,1}]=R_{4,1}^2.
\end{equation}
Actually, this relation determines the entire structure of all
algebraic invariants related to the dB-equation.

\subsection{$\mathcal{R}\circ\mathcal{H}$}
\label{subsec:rh}

We computed the following compositions:
\begin{align*}
  R_{4,1}\circ H_{0,1} =& \frac{1}{2}H_{4,2},\\
  R_{4,1}\circ H_{4,1} =& -H_{8,1}+H_{8,2}+2H_{8,3},\\
  R_{4,1}\circ H_{4,2} =& H_{8,1},\\[1mm]
  R_{4,2}\circ H_{0,1} =& \frac{1}{4} H_{4,1}-\frac{1}{4}H_{4,2},\\
  R_{4,2}\circ H_{4,1} =& -\frac{1}{2}H_{8,2}-\frac{1}{2}H_{8,3},\\
  R_{4,2}\circ H_{4,2} =& H_{8,1}+\frac{1}{2}H_{8,2}+
  \frac{1}H_{8,3},\\[1mm]
  R_{8,1}\circ H_{0,1} =& 8H_{8,1}-4H_{8,2}-8H_{8,3},\\
  R_{8,2}\circ H_{0,1} =& -\frac{18}{7} H_{8,1}+\frac{12}{7}H_{8,2}+
  \frac{24}{7}H_{8,3},\\
  R_{8,3}\circ H_{0,1} =& \frac{6}{13}H_{8,1}-\frac{7}{13}H_{8,2}-
  \frac{12}{13}H_{8,3}.
\end{align*}
etc. These relations imply, in particular, the equality
\begin{equation*}
  (3\mathcal{R}_{4,1}+4\mathcal{R}_{4,2})\circ\mathcal{H}_{0,1}=
  \mathcal{H}_{4,1}+\frac{1}{2}\mathcal{H}_{4,2}.
\end{equation*}
Denoting~$3\mathcal{R}_{4,1}+4\mathcal{R}_{4,2}$ by~$\mathcal{R}$, we
have
\begin{equation}
  \label{eq:bouss-new:69}
  \mathcal{R}\circ\mathcal{H}_1=\mathcal{H}_2,
\end{equation}
where~$\mathcal{H}_1$ and~$\mathcal{H}_2$ are the Hamiltonian
structures introduced in Subsection~\ref{subsec:hamilt-struct}.

\begin{proposition}
  \label{prop:Ham-ser}
  The operator
  \begin{equation}
    \label{eq:bouss-new:70}
    \mathcal{R}=3\mathcal{R}_{4,1}+4\mathcal{R}_{4,2}
  \end{equation}
  satisfies the Nijenhuis conditions \textup{(}see
  Remark~\ref{rem:recurs-oper-hamilt-3}\textup{)} and generates and
  infinite family of compatible Hamiltonian structures
  \begin{equation*}
    \mathcal{H}_n=\mathcal{R}^{n-1}\circ\mathcal{H}_1
  \end{equation*}
  for the dB-equation.
\end{proposition}

\subsection{$\mathcal{S}\circ\mathcal{R}$}
\label{subsec:sr}

These compositions look as follows
\begin {align*}
  S_{0,1}\circ R_{4,1} &= \frac{1}{4} S_{4,1} +\frac{7}{12} S_{4,2},\\
  S_{0,1}\circ R_{4,2} &= -\frac{1}{16} S_{4,1},\\
  S_{0,1}\circ R_{8,1} &= \frac{11}{2} S_{8,1} +\frac{1}{4} S_{8,2},\\
  S_{0,1}\circ R_{8,2} &= -\frac{33}{28} S_{8,1}+\frac{9}{28} S_{8,3},\\
  S_{0,1}\circ R_{8,3} &= \frac{11}{104} S_{8,1},\\[1mm]
  S_{4,1}\circ R_{4,1} &= -\frac{11}{2} S_{8,1} -\frac{3}{4} S_{8,2}
  -3S_{8,3},\\
  S_{4,1}\circ R_{4,2} &= \frac{11}{4} S_{8,1} +\frac{3}{16} S_{8,2},\\[1mm]
  S_{4,2}\circ R_{4,1} &= \frac{99}{28} S_{8,1}
  +\frac{3}{7} S_{8,2}+\frac{45}{28} S_{8,3},\\
  S_{4,2}\circ R_{4,2} &= -\frac{99}{56} S_{8,1} -\frac{3}{28}S_{8,2},
\end{align*}
etc.

\subsection{$\mathcal{H}\circ\mathcal{S}$}
\label{subsec:hs}
These compositions look as follows
\begin {align*}
  H_{0,1}\circ S_{0,1} =& \frac{1}{2} R_{0,1},\\
  H_{0,1}\circ S_{4,1}  =& -8R_{4,2},\\
  H_{0,1}\circ S_{4,2}  =& \frac{6}{7} R_{4,1}+\frac{24}{7}R_{4,2},\\
  H_{0,1}\circ S_{8,1}  =& \frac{52}{11} R_{8,3},\\
  H_{0,1}\circ S_{8,2}  =& 2R_{8,1}-104 R_{8,3},\\
  H_{0,1}\circ S_{8,3} =& R_{8,1}+\frac{14}{9}R_{8,2}+
  \frac{52}{3} R_{8,3},\\[1mm]
  H_{4,1}\circ S_{0,1} =& R_{4,1}+2R_{4,2},\\
  H_{4,1}\circ S_{4,1}  =& \frac{5}{2} R_{8,1}-26 R_{8,3},\\
  H_{4,1}\circ S_{4,2}  =& 3R_{8,2}+ \frac{78}{7} R_{8,3},\\[1mm]
  H_{4,2}\circ S_{0,1} =& R_{4,1},\\
  H_{4,2}\circ S_{4,1}  =& R_{8,1},\\
  H_{4,2}\circ S_{4,2}  =& R_{8,1}+R_{8,2},\\[1mm]
  H_{8,1}\circ S_{0,1} =& \frac{1}{4} R_{8,1}+\frac{7}{12}R_{8,2},\\
  H_{8,2}\circ S_{0,1} =& -\frac{3}{4}
  R_{8,1}-\frac{7}{2}R_{8,2}-\frac{13}{2} R_{8,3},\\
  H_{8,3}\circ S_{0,1} =& \frac{9}{16}R_{8,1}+\frac{7}{3}R_{8,2}+
  \frac{13}{4}R_{8,3},
\end{align*}
etc.

\subsection{$\mathcal{S}\circ\mathcal{H}$}
\label{subsec:sh}
Similar to the identities of the previous subsection, we have
\begin {align*}
  S_{0,1}\circ H_{0,1} =& \frac{1}{2} R_{0,1}^*,\\
  S_{0,1}\circ H_{4,1} =& -R_{4,1}^*+2R_{4,2}^*,\\
  S_{0,1}\circ H_{4,2} =& R_{4,1}^*,\\
  S_{0,1}\circ H_{8,1} =& \frac{1}{4} R_{8,1}^*+\frac{7}{12}R_{8,2}^*,\\
  S_{0,1}\circ H_{8,2} =& -\frac{1}{4} R_{8,1}^*-\frac{13}{2} R_{8,3}^*,\\
  S_{0,1}\circ H_{8,3} =& \frac{1}{16} R_{8,1}^*
  +\frac{13}{4}R_{8,3}^*,\\[1mm]
  S_{4,1}\circ H_{0,1} =&  8R_{4,1}^*-8R_{4,2}^*,\\
  S_{4,1}\circ H_{4,1} =&  -\frac{3}{2} R_{8,1}^*-26R_{8,3}^*,\\
  S_{4,1}\circ H_{4,2} =&  R_{8,1}^*,\\[1mm]  
  S_{4,2}\circ H_{0,1} =&  -\frac{18}{7} R_{4,1}^*
  +\frac{24}{7}R_{4,2}^*,\\
  S_{4,2}\circ H_{4,1} =&  -R_{8,2}^*+\frac{78}{7}R_{8,3}^*,\\
  S_{4,2}\circ H_{4,2} =&  R_{8,2}^*,\\[1mm]
  S_{8,1}\circ H_{0,1} =& \frac{20}{11}R_{8,1}^*+\frac{28}{11}R_{8,2}^*
  + \frac{52}{11} R_{8,3}^*,\\
  S_{8,2}\circ H_{0,1} =&  -30R_{8,1}^*-\frac{112}{3}R_{8,2}^*
  -104R_{8,3}^*,\\
  S_{8,3}\circ H_{0,1} =& 4R_{8,1}^*+\frac{14}{3}R_{8,2}^*+
  \frac{52}{3} R_{8,3}^*
\end{align*}
etc.

\subsection{$\mathcal{H}\circ\mathcal{R}^*$}
\label{subsec:hrstar}

We computed the following compositions of this type
\begin {align*}
  H_{0,1}\circ R_{4,1}^* &= \frac{1}{2} H_{4,2},\\
  H_{0,1}\circ R_{4,2}^* &= \frac{1}{4} H_{4,1} +\frac{1}{4} H_{4,2},\\
  H_{0,1}\circ R_{8,1}^* &= -4H_{8,2}-8H_{8,3},\\
  H_{0,1}\circ R_{8,2}^* &= \frac{6}{7} H_{8,1} +\frac{12}{7} H_{8,2}
  +\frac{24}{7} H_{8,3},\\
  H_{0,1}\circ R_{8,3}^* &= \frac{1}{13} H_{8,2}+\frac{4}{13} H_{8,3},\\[1mm]
  H_{4,1}\circ R_{4,1}^* &= H_{8,1} +H_{8,2}+H_{8,3},\\
  H_{4,1}\circ R_{4,2}^* &= \frac{1}{2} H_{8,2}+\frac{3}{2} H_{8,3},\\[1mm]
  H_{4,2}\circ R_{4,1}^* &= H_{8,1} ,\\
  H_{4,2}\circ R_{4,2}^* &= \frac{1}{2} H_{8,2}+H_{8,3},
\end{align*}
which are in a sense dual to the results of
Subsection~\ref{subsec:hrstar}.

\subsection{$\mathcal{R}^*\circ\mathcal{S}$}
\label{subsec:rstars}

Computing compositions of recursion operators with
operators~$\mathcal{S}$ we obtain
\begin{align*}
  R_{4,1}^*\circ S_{0,1} =& \frac{ 1}{4} S_{4,1}+\frac{7}{12}S_{4,2},\\
  R_{4,1}^*\circ S_{4,1} =& \frac{11}{2} S_{8,1}+\frac{1}{4}S_{8,2},\\
  R_{4,1}^*\circ S_{4,2} =& -\frac{33}{28} S_{8,1}+
  \frac{9}{28}S_{8,3},\\[1mm]
  R_{4,2}^*\circ S_{0,1} =& \frac{3}{16} S_{4,1}+\frac{7}{12} S_{4,2},\\
  R_{4,2}^*\circ S_{4,1} =& \frac{33}{4} S_{8,1}+\frac{7}{16}S_{8,2},\\
  R_{4,2}^*\circ S_{4,2} =& -\frac{99}{56} S_{8,1}+
  \frac{9}{14}S_{8,3},\\[1mm]
  R_{8,1}^*\circ S_{0,1} =& -\frac{11}{2} S_{8,1}-\frac{3}{4}S_{8,2}
  -\frac{3}S_{8,3},\\
  R_{8,2}^*\circ S_{0,1} =& \frac{99}{28} S_{8,1}+\frac{3}{7}S_{8,2}
  + \frac{45}{28}S_{8,3},\\
  R_{8,3}^*\circ S_{0,1} =& \frac{33}{104} S_{8,1}+\frac{3}{52}S_{8,2}
  + \frac{15}{52}S_{8,3},
\end{align*}
etc. In particular, we have the relation
\begin{equation*}
  (4\mathcal{R}_{4,2}^*-\mathcal{R}_{4,1}^*)\circ\mathcal{S}_1=
  \frac{1}{4}\mathcal{S}_2,
\end{equation*}
where
\begin{equation*}
  \mathcal{S}_1=\mathcal{S}_{0,1},\quad
  \mathcal{S}_2=2\mathcal{S}_{4,1}+7\mathcal{S}_{4,2}
\end{equation*}
are the symplectic structures introduced in
Subsection~\ref{subsec:sympl-struct}. Moreover, we have the following
result:

\begin{proposition}
  The operator
  \begin{equation}\label{eq:bouss-new:73}
    \mathcal{R}^*=4(4\mathcal{R}_{4,2}^*-\mathcal{R}_{4,1}^*)
  \end{equation}
  generates an infinite family of symplectic structures for the
  dB-equation by
  \begin{equation*}
    \mathcal{S}_n=(\mathcal{R}^*)^{n-1}\circ\mathcal{S}_1.
  \end{equation*}
\end{proposition}

\subsection{$\mathcal{R}^*\circ\mathcal{R}^*$}
\label{subsec:rstarrstar}

The first relations we obtained are
\begin {align*}
  R_{4,1}^*\circ R_{4,1}^* &=\frac{1}{4}R_{8,1}^*+\frac{7}{12}R_{8,2}^*,\\
  R_{4,1}^*\circ R_{4,2}^* &=-\frac{1}{16}R_{8,1}^*,\\
  R_{4,2}^*\circ R_{4,1}^* &=\frac{3}{16}R_{8,1}^*+\frac{7}{12}R_{8,2}^*,\\
  R_{4,2}^*\circ R_{4,2}^*
  &=-\frac{3}{64}R_{8,1}^*+\frac{13}{16}R_{8,3}^*.
\end{align*}
Hence, the relation
\begin{equation*}
  R_{4,2}^*\circ R_{4,1}^*=R_{4,1}^*\circ R_{4,2}^*+R_{4,1}^*\circ R_{4,1}^* 
\end{equation*}
takes place, or
\begin{equation}
  \label{eq:bouss-new:72}
  [R_{4,2}^*,R_{4,1}^*]=R_{4,1}^*\circ R_{4,1}^*
\end{equation}
that is similar to the one we had for the operators~$\mathcal{R}$ (see
equation~\eqref{eq:bouss-new:68}).

\subsection{Action of $\mathcal{R}$ on symmetries}
\label{subsec:action-symm}

\subsubsection{Action on local symmetries}
\label{ssubsec:acti-local-symm}
The action of the operator~$R_{4,1}$ is as follows
\begin{align*}
  R_{4,1}(\phi_{1})&= \frac{1}{2}\phi_{5},\ &R_{4,1}(\phi_{2}) &=
  \frac{1}{4}\phi_{6},\ 
  &R_{4,1}(\phi_{3})  &= \frac{1}{12}\phi_{7},\\
  R_{4,1}(\phi_{5}) &= \frac{1}{8}\phi_{9},\ &R_{4,1}(\phi_{6}) &=
  \frac{1}{24}\phi_{10},\ 
  &R_{4,1}(\phi_{7})  &= \frac{1}{4}\phi_{11},\\
  R_{4,1}(\phi_{9}) &= \frac{1}{60}\phi_{13},\ &R_{4,1}(\phi_{10}) &=
  \frac{1}{4}\phi_{14},\ &R_{4,1}(\phi_{11}) &= \frac{1}{28}\phi_{15}
\end{align*}
and is related with the action of~$R_{4,2}$ in the following way
\begin{equation}\label{eq:bouss-new:71}
R_{4,1}(\phi_{i})  = \frac{4}{i+1}R_{4,2}(\phi_{i}).
\end{equation}

\subsubsection{Action on $(x,t)$-independent nonlocal symmetries}
\label{ssubsec:action-x-t-ind}
Here we have
\begin{align*}
  R_{4,1}(\phi_{4,1}) &=-\frac{1}{104}\phi_{8,2},\\
  R_{4,1}(\phi_{8,1}) &=385\phi_{12,1}+21\phi_{12,2},\\
  R_{4,1}(\phi_{8,2})
  &=1430\phi_{12,1}+156\phi_{12,2}+234\phi_{12,3}
\end{align*}
and
\begin{align*}
  R_{4,2}(\phi_{4,1}) &=\frac{1}{56}\phi_{8,1}-\frac{1}{104}\phi_{8,2}\\
  R_{4,2}(\phi_{8,1}) &=\frac{1155}{2}\phi_{12,1}+\frac{147}{4}\phi_{12,2}\\
  R_{4,2}(\phi_{8,2})
  &=2145\phi_{12,1}+273\phi_{12,2}+468\phi_{12,3}
\end{align*}
etc.

\subsubsection{Action on $(x,t)$-dependent nonlocal symmetries}
\label{ssubsec:action-x-t-dep}
We have
\begin{align*}
  R_{4,1}(\ol{\phi}_{0,1})&= \frac{1}{2}\ol{\phi}_{4,1}\\
  R_{4,1}(\ol{\phi}_{0,2})&= \frac{1}{4}\ol{\phi}_{4,2}\\
  R_{4,1}(\ol{\phi}_{4,1})&= -\frac{1}{208}\phi_{8,2}
  +\frac{1}{8}\ol{\phi}_{8,1}\\
  R_{4,1}(\ol{\phi}_{4,2})&= \frac{1}{156}\phi_{8,2}
  +\frac{1}{24}\ol{\phi}_{8,2}\\
  R_{4,1}(\ol{\phi}_{8,1})&= 2\phi_{12,2}+\frac{55}{2}\phi_{12,1}
  +\frac{1}{60}\ol{\phi}_{12,1}\\
  R_{4,1}(\ol{\phi}_{8,2})&=
  -6\phi_{12,2}-88\phi_{12,1}+\frac{1}{4}\ol{\phi}_{12,2}
\end{align*}
and
\begin{align*}
  R_{4,2}(\ol{\phi}_{0,1})&=-\frac{1}{8}\phi_{4,1}
  +\frac{1}{4}\ol{\phi}_{4,1},\\
  R_{4,2}(\ol{\phi}_{0,2})&=\frac{1}{8}\phi_{4,1}
  +\frac{3}{16}\ol{\phi}_{4,2},\\
  R_{4,2}(\ol{\phi}_{4,1})&= -\frac{1}{208}\phi_{8,2}
  +\frac{3}{16}\ol{\phi}_{8,1},\\
  R_{4,2}(\ol{\phi}_{4,2})&= \frac{1}{156}\phi_{8,2}
  +\frac{7}{96}\ol{\phi}_{8,2},\\
  R_{4,2}(\ol{\phi}_{8,1})&= \frac{7}{2}\phi_{12,2}
  +\frac{165}{4}\phi_{12,1}+\frac{1}{24}\ol{\phi}_{12,1},\\
  R_{4,2}(\ol{\phi}_{8,2})&= -\frac{21}{2}\phi_{12,2}
  -132\phi_{12,1}+\frac{11}{16}\ol{\phi}_{12,2}.
\end{align*}

\subsection{Action of $\mathcal{R}^*$ on cosymmetries}
\label{subsec:action-cosymm}

In a similar way, we have the following actions of~$\mathcal{R}^*$:

\subsubsection{Action on local cosymmetries}
\label{ssubsec:acti-local-cosymm}

\begin {align*}
R_{4,1}^*(\psi_{2}) &= \frac{1}{2} \psi_{6},
&R_{4,1}^*(\psi_{3}) &= \frac{1}{2} \psi_{7},
&R_{4,1}^*(\psi_{4}) &= \frac{7}{12} \psi_{8},\\
R_{4,1}^*(\psi_{6}) &= \frac{11}{8} \psi_{10},
&R_{4,1}^*(\psi_{7}) &= \frac{1}{4} \psi_{11},
&R_{4,1}^*(\psi_{8}) &= \frac{1}{56} \psi_{12},\\
R_{4,1}^*(\psi_{10})&= \frac{91}{330} \psi_{14},
&R_{4,1}^*(\psi_{11})&= \frac{1}{8} \psi_{15},
&R_{4,1}^*(\psi_{12})&= \frac{57}{14} \psi_{16}
\end{align*}
and
\begin {equation}
R_{4,2}^*(\psi_{i})=\frac{i}{4}R_{4,1}^*(\psi_{i}).
\end{equation}

\subsubsection{Action on $(x,t)$-independent nonlocal cosymmetries}
\label{ssubsec:action-x-t-co}

\begin {align*}
  R^*_{4,1}( \psi_{9,1})&= \frac{3}{14} \psi_{13,1}
  +\frac{9}{28} \psi_{13,2},\\
  R^*_{4,1}( \psi_{13,1})&= \frac{91}{18} \psi_{17,1}+
  \frac{1}{8} \psi_{17,2},\\
  R^*_{4,1}( \psi_{13,2})&= -\frac{91}{54} \psi_{17,1}
  +\frac{19}{84} \psi_{17,3},\\[1mm]
  R^*_{4,2}( \psi_{9,1})&= \frac{3}{8} \psi_{13,1}
  +\frac{9}{14} \psi_{13,2},\\
  R^*_{4,2}( \psi_{13,1})&= \frac{455}{36} \psi_{17,1}
  +\frac{11}{32} \psi_{17,2},\\
  R^*_{4,2}( \psi_{13,2})&= -\frac{455}{108} \psi_{17,1}
  +\frac{19}{28} \psi_{17,3}.
\end{align*}

\subsubsection{Action on $(x,t)$-dependent nonlocal cosymmetries}
\label{ssubsec:action-x-t-dep-co}

\begin {align*}
  R_{4,1}^*\circ\ol{\psi}_{1,1}&=\frac{1}{2}\ol{\psi}_{5,1}
  +\frac{1}{2}\ol{\psi}_{5,2},\\
  R_{4,1}^*\circ\ol{\psi}_{5,1}&=-\frac{7}{2}\psi_{9,1}
  +\frac{11}{8}\ol{\psi}_{9,1},\\
  R_{4,1}^*\circ\ol{\psi}_{5,2}&=\frac{7}{3}\psi_{9,1}
  +\frac{1}{4}\ol{\psi}_{9,2},\\
  R_{4,1}^*\circ\ol{\psi}_{9,1}&=\frac{2}{11}\psi_{13,1}
  +\frac{91}{330}\ol{\psi}_{13,1},\\
  R_{4,1}^*\circ\ol{\psi}_{9,2}&=-\frac{1}{2}\psi_{13,1}
  +\frac{1}{8}\ol{\psi}_{13,2},\\[1mm]
  R_{4,2}^*\circ\ol{\psi}_{1,1}&=\frac{1}{4}\ol{\psi}_{5,1}
  +\frac{3}{8}\ol{\psi}_{5,2},\\
  R_{4,2}^*\circ\ol{\psi}_{5,1}&=-\frac{7}{2}\psi_{9,1}
  +\frac{33}{16}\ol{\psi}_{9,1},\\
  R_{4,2}^*\circ\ol{\psi}_{5,2}&=\frac{7}{3}\psi_{9,1}
  +\frac{7}{16}\ol{\psi}_{9,2},\\
  R_{4,2}^*\circ\ol{\psi}_{9,1}&=\frac{7}{22}\psi_{13,1}
  +\frac{91}{132}\ol{\psi}_{13,1},\\
  R_{4,2}^*\circ\ol{\psi}_{9,2}&=-\frac{7}{8}\psi_{13,1}
  +\frac{11}{32}\ol{\psi}_{13,2}.
\end{align*}

\subsection{Commutators}
\label{subsec:commutators}

\subsubsection{$[\bullet,\phi_i]$-commutators}
\label{ssubsec:phi_i-comm}

First of all we have the relation
\[
[\phi_{i},\phi_{j}]=0,
\]
and besides these relations we obtain:
\begin{align*}
  [\phi_{4,1},\phi_{1}]  &= 0,
  &[\phi_{8,1},\phi_{1}] &= 0,
  &[\phi_{8,2},\phi_{1}] &= 0,\\
  [\phi_{4,1},\phi_{2}]  &= 0,
  &[\phi_{8,1},\phi_{2}] &= 0,
  &[\phi_{8,2},\phi_{2}] &= 0,\\
  [\phi_{4,1},\phi_{3}]  &= \frac{1}{3}\phi_{7},
  &[\phi_{8,1},\phi_{3}] &= \frac{14}{3}\phi_{11},
  &[\phi_{8,2},\phi_{3}] &= -\frac{26}{3}\phi_{11},\\
  [\phi_{4,1},\phi_{5}]  &= 3\phi_{9}
  &[\phi_{8,1},\phi_{5}] &= \frac{21}{5}\phi_{13},
  &[\phi_{8,2},\phi_{5}] &= -\frac{26}{5}\phi_{13},\\
  [\phi_{4,1},\phi_{6}]  &= \frac{5}{3}\phi_{10},
  &[\phi_{8,1},\phi_{6}] &= \frac{245}{6}\phi_{14},
  &[\phi_{8,2},\phi_{6}] &= -\frac{130}{3}\phi_{14},\\
  [\phi_{4,1},\phi_{7}]  &= 15\phi_{11},
  &[\phi_{8,1},\phi_{7}] &= 60\phi_{15},
  &[\phi_{8,2},\phi_{7}] &= -\frac{390}{7}\phi_{15},\\
  [\phi_{4,1},\phi_{9}]  &= \frac{28}{15}\phi_{13},
  &                      &
  &                      &\\
  [\phi_{4,1},\phi_{10}] &= 36\phi_{14},
  &                      &
  &                      &\\
  [\phi_{4,1},\phi_{11}] &= \frac{45}{7}\phi_{15},
  &                      &
  &                      &
\end{align*}
and
\begin{align*}
  [\ol{\phi}_{0,1},\phi_{1}] &= \phi_{1},
  &[\ol{\phi}_{0,2},\phi_{1}]&= 0,\\
  [\ol{\phi}_{0,1},\phi_{2}] &= 0,
  &[\ol{\phi}_{0,2},\phi_{2}]&= \phi_{2},\\
  [\ol{\phi}_{0,1},\phi_{3}] &= -\phi_{3},
  &[\ol{\phi}_{0,2},\phi_{3}]&= 2\phi_{3},\\
  [\ol{\phi}_{0,1},\phi_{5}] &= -3\phi_{5},
  &[\ol{\phi}_{0,2},\phi_{5}]&= 4\phi_{5},\\
  [\ol{\phi}_{0,1},\phi_{6}] &= -4\phi_{6},
  &[\ol{\phi}_{0,2},\phi_{6}]&= 5\phi_{6},\\
  [\ol{\phi}_{0,1},\phi_{7}] &= -5\phi_{7},
  &[\ol{\phi}_{0,2},\phi_{7}]&= 6\phi_{7},\\[1mm]
  [\ol{\phi}_{4,1},\phi_{1}] &= \phi_{5},
  &[\ol{\phi}_{4,2},\phi_{1}]&= 0,\\
  [\ol{\phi}_{4,1},\phi_{2}] &= 0,
  &[\ol{\phi}_{4,2},\phi_{2}]&= \phi_{6},\\
  [\ol{\phi}_{4,1},\phi_{3}] &= -\frac{1}{6} \phi_{7},
  &[\ol{\phi}_{4,2},\phi_{3}]&= \frac{2}{3} \phi_{7},\\
  [\ol{\phi}_{4,1},\phi_{5}] &= -\frac{3}{4} \phi_{9},
  &[\ol{\phi}_{4,2},\phi_{5}]&= 2\phi_{9},\\
  [\ol{\phi}_{4,1},\phi_{6}] &= -\frac{1}{3} \phi_{10},
  &[\ol{\phi}_{4,2},\phi_{6}]&= \frac{5}{6} \phi_{1},\\
  [\ol{\phi}_{4,1},\phi_{7}] &= -\frac{5}{2} \phi_{11},
  &[\ol{\phi}_{4,2},\phi_{7}]&= \frac{6} \phi_{11},\\[1mm]
  [\ol{\phi}_{8,1},\phi_{1}] &= \phi_{9},
  &[\ol{\phi}_{8,2},\phi_{1}]&= 0,\\
  [\ol{\phi}_{8,1},\phi_{2}] &= 0,
  &[\ol{\phi}_{8,2},\phi_{2}]&= \phi_{10},\\
  [\ol{\phi}_{8,1},\phi_{3}] &= -\frac{2}{3} \phi_{11},
  &[\ol{\phi}_{8,2},\phi_{3}]&= \frac{16}{3} \phi_{11},\\
  [\ol{\phi}_{8,1},\phi_{5}] &= -\frac{3}{10} \phi_{13},
  &[\ol{\phi}_{8,2},\phi_{5}]&= \frac{8}{5} \phi_{13},\\
  [\ol{\phi}_{8,1},\phi_{6}] &= -\frac{7}{3} \phi_{14},
  &[\ol{\phi}_{8,2},\phi_{6}]&= \frac{35}{3} \phi_{14},\\
  [\ol{\phi}_{8,1},\phi_{7}] &= -\frac{20}{7} \phi_{15},
  &[\ol{\phi}_{8,2},\phi_{7}]&= \frac{96}{7} \phi_{15}.
\end{align*}

\subsubsection{$[\bullet,\phi_{i,j}]$-commutators}
\label{ssubsec:bullet-phi_i-j}

We computed the following ones
\begin{align*}
[\phi_{4,1},\phi_{8,1}] &= 9240\phi_{12,1}+840\phi_{12,2}+504\phi_{12,3},\\
[\phi_{4,1},\phi_{8,2}] &= 34320\phi_{12,1}+6240\phi_{12,2}+13104\phi_{12,3}
\end{align*}
and
\begin{align*}
  [\ol{\phi}_{0,1},\phi_{4,1}]  &= -4\phi_{4,1},\\
  [\ol{\phi}_{0,1},\phi_{8,1}]  &= -8\phi_{8,1},\\
  [\ol{\phi}_{0,1},\phi_{8,2}]  &= -8\phi_{8,2},\\
  [\ol{\phi}_{0,2},\phi_{4,1}]  &= 4\phi_{4,1},\\
  [\ol{\phi}_{0,2},\phi_{8,1}]  &= 8\phi_{8,1},\\
  [\ol{\phi}_{0,2},\phi_{8,2}]  &= 8\phi_{8,2},\\
  [\ol{\phi}_{4,1},\phi_{4,1}]  &= \frac{1}{26}\phi_{8,2}
  -3\ol{\phi}_{8,1},\\
  [\ol{\phi}_{4,1},\phi_{8,1}]  &= 252\phi_{12,3}-168\phi_{12,2}
  -2310\phi_{12,1}-\frac{21}{5}\ol{\phi}_{12,1},\\
  [\ol{\phi}_{4,1},\phi_{8,2}]  &= -2808\phi_{12,3}-1248\phi_{12,2}
  -8580\phi_{12,1}+\frac{26}{5}\ol{\phi}_{12,1},\\
  [\ol{\phi}_{4,2},\phi_{4,1}]  &= -\frac{4}{39}\phi_{8,2}
  -\frac{5}{3}\ol{\phi}_{8,2},\\
  [\ol{\phi}_{4,2},\phi_{8,1}]  &= -336\phi_{12,3}+420\phi_{12,2}
  +6160\phi_{12,1}-\frac{245}{6}\ol{\phi}_{12,2},\\
  [\ol{\phi}_{4,2},\phi_{8,2}]  &= 6240\phi_{12,3}+3120\phi_{12,2}
  +22880\phi_{12,1}+\frac{130}{3}\ol{\phi}_{12,2},\\
  [\ol{\phi}_{8,1},\phi_{4,1}]  &= 72\phi_{12,3}-80\phi_{12,2}-
  880\phi_{12,1}-\frac{28}{15}\ol{\phi}_{12,1},\\
  [\ol{\phi}_{8,2},\phi_{4,1}] &=
  -576\phi_{12,3}+96\phi_{12,2}+2112\phi_{12,1}-36\ol{\phi}_{12,2}.
\end{align*}

\subsubsection{$[\bullet,\bar\phi_{i,j}]$-commutators}
\label{ssubsec:bullet-barphi_i-j}
We computed the following commutators of this type
\begin{align*}
  [\ol{\phi}_{0,1},\ol{\phi}_{0,2}]  &= 0,\\
  [\ol{\phi}_{0,1},\ol{\phi}_{4,1}]  &= -4\ol{\phi}_{4,1},\\
  [\ol{\phi}_{0,1},\ol{\phi}_{4,2}]  &= -4\ol{\phi}_{4,2},\\
  [\ol{\phi}_{0,1},\ol{\phi}_{8,1}]  &= -8\ol{\phi}_{8,1},\\
  [\ol{\phi}_{0,1},\ol{\phi}_{8,2}]  &= -8\ol{\phi}_{8,2},\\
  [\ol{\phi}_{0,2},\ol{\phi}_{4,1}]  &= 4\ol{\phi}_{4,1},\\
  [\ol{\phi}_{0,2},\ol{\phi}_{4,2}]  &= 4\ol{\phi}_{4,2},\\
  [\ol{\phi}_{0,2},\ol{\phi}_{8,1}]  &= 8\ol{\phi}_{8,1},\\
  [\ol{\phi}_{0,2},\ol{\phi}_{8,2}]  &= 8\ol{\phi}_{8,2},\\
  [\ol{\phi}_{4,1},\ol{\phi}_{4,2}]  &= \frac{1}{39}\phi_{8,2}
  -2\ol{\phi}_{8,1}-\frac{1}{3}\ol{\phi}_{8,2},\\
  [\ol{\phi}_{4,1},\ol{\phi}_{8,1}]  &= -36\phi_{12,3}-16\phi_{12,2}
  -110\phi_{12,1}+\frac{1}{15}\ol{\phi}_{12,1},\\
  [\ol{\phi}_{4,1},\ol{\phi}_{8,2}]  &= 288\phi_{12,3}+96\phi_{12,2}
  +528\phi_{12,1}-\frac{8}{5}\ol{\phi}_{12,1}-4\ol{\phi}_{12,2},\\
  [\ol{\phi}_{4,2},\ol{\phi}_{8,1}]  &= 48\phi_{12,3}+40\phi_{12,2}
  +352\phi_{12,1}+\frac{8}{15}\ol{\phi}_{12,1}
  +\frac{7}{3}\ol{\phi}_{12,2},\\
  [\ol{\phi}_{4,2},\ol{\phi}_{8, 2}] &= -384\phi_{12,3}
  -192\phi_{12,2}-1408\phi_{12,1}-\frac{8}{3}\ol{\phi}_{12,2}.
\end{align*}

\section{Conclusions}
\label{sec:discussion}

We studied the dispersionless Boussinesq equation
\begin{align*}
  w_t&=u_x,\\
  u_t&=ww_x+v_x,\\
  v_t&=-uw_x-3wu_x
\end{align*}
and found out that it possesses three families of commuting local
symmetries,~$\Phi_1$, $\Phi_2$ and~$\Phi_3$. The seed symmetries for
these families are~$\phi_1$, $\phi_2$ and~$\phi_3$ presented in
Subsection~\ref{ssubsec:x-t-independent}. In addition, the equation
has a family~$\Phi_4$ of nonlocal $(x,t)$-independent symmetries
generated by the symmetry~$\phi_{4,1}$
(Subsection~\ref{ssubsec:nonlocal-symmetries}) and two
families,~$\bar{\Phi}_1$ and~$\bar{\Phi}_2$, of nonlocal
$(x,t)$-dependent symmetries generated by $\bar{\phi}_{0,1}$
and~$\bar{\phi}_{0,2}$ (Subsection~\ref{ssubsec:x-t-dependent}). All
nonlocal symmetries act, by the commutator, as hereditary symmetries
for the local ones.

We also constructed an infinite algebra~$\Rec(\mathcal{E})$ of
recursion operators for symmetries. This is an associative
noncommutative algebra with two generators~$\mathcal{R}_{4,1}$
and~$\mathcal{R}_{4,2}$ and one relation (see
Subsection~\ref{subsec:recurs-oper-symm} and
equation~\eqref{eq:bouss-new:68}). The operators produce the above
mentioned families from the corresponding seed symmetries. This
algebra contains an Abelian subalgebra of operators satisfying the
Nijenhuis condition. The subalgebra is generated by
operator~\eqref{eq:bouss-new:70}. Existence of this operator is the
reason for commutativity of the families~$\Phi_1$, $\Phi_2$
and~$\Phi_3$.

In complete parallel, one observes five families of cosymmetries and
an algebra of the corresponding recursion operator that generate these
families. This similarity is due to existence of
operators~$\mathcal{H}\colon\sym^*(\mathcal{E})\to\sym(\mathcal{E})$
and~$\mathcal{S}\colon\sym(\mathcal{E})\to\sym^*(\mathcal{E})$ that
relate symmetries and cosymmetries to each other and are described in
Subsections~\ref{subsec:hamilt-struct}
and~\ref{subsec:sympl-struct}, resp.

The operators~$\mathcal{H}$ form a left module over the
algebra~$\Rec(\mathcal{E})$ with one generator~$\mathcal{H}_{0,1}$.
This operator is a Hamiltonian structure for the dB-equation and by
the action of recursion operator~\eqref{eq:bouss-new:70} one obtains
an infinite family of compatible Hamiltonian structures of which the
first two are local. In a similar way, the operators~$\mathcal{S}$
form a left module over the algebra of recursion operators for
cosymmetries, while the recursion operator given by
equation~\eqref{eq:bouss-new:73} generates an infinite family of
symplectic structures, all of them being nonlocal. 

On the other hand, the dB-equation~$\mathcal{E}$ admits the
potential~$y$ defined by
\begin{equation*}
  y_x=w,\qquad y_t=u
\end{equation*}
identical to the nonlocal variable~$p_1$ (see
Subsection~\ref{subsec:nonlocal-variables}). The corresponding
covering equation~$\mathcal{E}_1$ is of the form
\begin{equation*}
  y_{tt}=y_xy_{xx}+v_t,\qquad v_t=-y_ty_{xx}-3y_xy_{xt}.
\end{equation*}
The potential~$z$ defined by
\begin{equation*}
  z_x=u,\qquad z_t=v+\frac{1}{2}w^2,
\end{equation*}
that corresponds to the nonlocal variable~$p_2$ leads to the covering
equation~$\mathcal{E}_2$
\begin{equation*}
  w_t=z_{xx},\qquad z_{tt}-2ww_t=-z_xw_x-3wz_{xx}.
\end{equation*}
Finally, the equation~$\mathcal{E}'$
\begin{equation*}
  y_{ttt}+2y_xy_{xxt}+y_ty_{xtt}+3y_{xx}y_{xt}=0.
\end{equation*}
This means that we have the following covering structure
\begin{equation*}
  \begin{CD}
    \mathcal{E}@<<< \mathcal{E}_1\\
    @AAA@VVV\\
    \mathcal{E}_2@>>>\mathcal{E}'
  \end{CD}
\end{equation*}
that relates equations~$\mathcal{E}$ and~$\mathcal{E}'$. From this
structure it follows that an auto-B\"{a}cklund transformation can be
constructed for the dB-equation. Existence of this transformation may
be closely related to the algebraic structures described in the
preceding sections.


\begin{thebibliography}{10}
  
\bibitem{KrasilshchikVinogradov:SCLDEqMP} A.~V. Bocharov, V.~N.
  Chetverikov, S.~V. Duzhin, N.~G. Khor{\cprime}kova, I.~S.
  Krasil{\cprime}shchik, A.~V. Samokhin, Yu.~N. Torkhov, A.~M.
  Verbovetsky, and A.~M. Vinogradov, \emph{Symmetries and conservation
    laws for differential equations of mathematical physics},
  Monograph, Amer. Math. Soc., 1999.
 
\bibitem{Rubtsov-Orlov} B.~Enriquez, A.~Orlov, and V.~Rubtsov,
  \emph{Higher Hamiltonian structures \textup{(}the $sl_2$
    case\textup{)}}, JETP Letters, \textbf{58} (1993) no.~8, 658-664.
    
\bibitem{Gessler} D.~M.~Gessler, \emph{On the Vinogradov
    $\mathcal{C}$-spectral sequence for determined systems of
    differential equations}, Diff.\ Geom.\ Appl. \textbf{7} (1997),
  303--324,
  \urlprefix\url{http://diffiety.ac.ru/preprint/98/09_98abs.htm}.

\bibitem{GumNut} H.~G\"umral, Y.~Nutku, \emph{Bi-Hamiltonian
    structures of D-Boussinesq and Benney--Lax equations},
  J.~Phys.~A:~Math.~Gen, \textbf{27}, (1994), 193--200.
  
\bibitem{IgoninVerbovetskyVitolo:FLVDOp} S.~Igonin, A.~Verbovetsky,
  and R.~Vitolo, \emph{On the formalism of local variational
    differential operators}, Memorandum 1641, Faculty of Mathematical
  Sciences, University of Twente, The Netherlands, 2002,
  \urlprefix\url{http://www.math.utwente.nl/publications/2002/1641abs.html}.
    
\bibitem{KerstenKrasilshchikVerbovetsky:HOpC} P.~Kersten,
  I.~Krasil{\cprime}shchik, and A.~Verbovetsky, \emph{Hamiltonian
    operators and $\ell^*$-coverings}, J.\ Geom.\ and Phys.,
  \textbf{50} (2004) 273--302, \eprint{math.DG/0304245}.
  
\bibitem{KerstenKrasilshchikVerbovetsky:N=2} P.~Kersten,
  I.~Krasil{\cprime}shchik, and A.~Verbovetsky,
  \emph{\textup{(}Non\textup{)}local Hamiltonian and symplectic
    structures, recursions, and hierarchies\textup{:} a new approach
    and applications to the $N=1$ supersymmetric KdV equation}, J.\ 
  Phys.\ A: Mathematical and General, \textbf{37} (2004) no.~18,
  5003--5019, \eprint{nlin.SI/0305026}.
  
\bibitem{KKV-AAM} P.~Kersten, I.~Krasil{\cprime}shchik, and
  A.~Verbovetsky, \emph{On the integrability conditions for some
    structures related to evolution differential equations}, Acta
  Appl.\ Math., \textbf{83} (2004) no.~1-2, 167--173,
  \eprint{math.DG/0310451}.
  
\bibitem{KKLS} I.~S. Krasil{\cprime}shchik, \emph{Algebras with flat
    connections and symmetries of differential equations}, in: Lie
  Groups and Lie Algebras: Their Representations, Generalizations and
  Applications, Kluwer Acad.\ Publ., Dordrecht, Boston, London, 1998,
  407-424.
  
\bibitem{KrasilshchikKersten:SROpCSDE} I.~S. Krasil{\cprime}shchik and
  P.~H.~M. Kersten, \emph{Symmetries and recursion operators for
    classical and supersymmetric differential equations}, Kluwer,
  2000.
  
\bibitem{KrasilshchikVinogradov:NTGDEqSCLBT} I.~S.
  Krasil{\cprime}shchik and A.~M. Vinogradov, \emph{Nonlocal trends in
    the geometry of differential equations: {S}ymmetries, conservation
    laws, and B\"{a}cklund transformations}, Acta Appl. Math.
  \textbf{15} (1989), 161--209.
  
\bibitem{KrasilshchikVerbovetsky:HMEqMP} J.~Krasil{\cprime}shchik and
  A.~M. Verbovetsky, \emph{Homological methods in equations of
    mathematical physics}, Advanced Texts in Mathematics, Open
  Education \& Sciences, Opava, 1998, \eprint{math.DG/9808130}.
    
\end{thebibliography}
\end{document}